\documentclass[twocolumn,english,superscriptaddress,nature]{revtex4}

\usepackage[utf8]{inputenc}
\usepackage{amsmath}
\usepackage{amsfonts} 
\usepackage{babel}
\usepackage{bm}
\usepackage{color}
\usepackage{dcolumn}
\usepackage{graphics}
\usepackage{graphicx}
\usepackage[rightcaption]{sidecap}
\usepackage{helvet}
\usepackage{longtable}
\usepackage{subfigure}
\usepackage{xspace}
\usepackage{xr}
\usepackage[colorlinks=true,linkcolor=blue,citecolor=blue,urlcolor=blue]{hyperref}
\usepackage[normalem]{ulem}

\setcitestyle{super}

\begin{document}

\title{Field-induced quasi-bound state within the two-magnon continuum of a\\square-lattice Heisenberg antiferromagnet}

\author{F.~Elson}
\affiliation{Department of Applied Physics, KTH Royal Institute of Technology, SE-106 91 Stockholm, Sweden}

\author{M.~Nayak}
\affiliation{Institute of Physics, Ecole Polytechnique F\'ed\'erale de Lausanne (EPFL), CH-1015 Lausanne, Switzerland}
\affiliation{Department of Physics and Astronomy, University of Tennessee, Knoxville, TN 37996, USA} 

\author{A.~A.~Eberharter}
\affiliation{PSI Center for Scientific Computing, Theory and Data, Paul Scherrer Institute, CH-5232 Villigen-PSI, Switzerland}

\author{M.~Skoulatos}
\affiliation{PSI Center for Neutron and Muon Sciences, Paul Scherrer Institute, CH-5232 Villigen-PSI, Switzerland}
\affiliation{Heinz Maier-Leibnitz Zentrum (MLZ) and Physics Department, Technical University of Munich, D-85748 Garching, Germany}

\author{S.~Ward}
\email[present address: Novo Nordisk A/S, Research and Early Development, 
Novo Nordisk Park 1, DK-2760  M{\aa}l\o{}v, Denmark]{}
\affiliation{PSI Center for Neutron and Muon Sciences, Paul Scherrer Institute, CH-5232 Villigen-PSI, Switzerland}

\author{U.~Stuhr}
\affiliation{PSI Center for Neutron and Muon Sciences, Paul Scherrer Institute, CH-5232 Villigen-PSI, Switzerland}

\author{N.~B.~Christensen}
\affiliation{Department of Physics, Technical University of Denmark, 2800 Kgs.~Lyngby, Denmark}

\author{D.~Voneshen}
\affiliation{ISIS Pulsed Neutron and Muon Source, Rutherford Appleton Laboratory, Didcot OX11 0QX, United Kingdom}
\affiliation{Department of Physics, Royal Holloway University of London, Egham TW20 0EX, United Kingdom}

\author{C.~Fiolka}
\affiliation{Department of Chemistry, Biochemistry and Pharmaceutical Sciences, University of Bern, Freiestrasse 3, CH-3012 Bern, Switzerland}

\author{K.~W.~Kr\"amer}
\affiliation{PSI Center for Neutron and Muon Sciences, Paul Scherrer Institute, CH-5232 Villigen-PSI, Switzerland}
\affiliation{Department of Chemistry, Biochemistry and Pharmaceutical Sciences, University of Bern, Freiestrasse 3, CH-3012 Bern, Switzerland}

\author{Ch.~R\"uegg}
\affiliation{Institute of Physics, Ecole Polytechnique F\'ed\'erale de Lausanne (EPFL), CH-1015 Lausanne, Switzerland}
\affiliation{Paul Scherrer Institute, CH-5232 Villigen-PSI, Switzerland}
\affiliation{Institute for Quantum Electronics, ETH Z\"urich, CH-8093 H\"onggerberg, Switzerland}
\affiliation{Department of Quantum Matter Physics, University of Geneva, CH-1211 Geneva, Switzerland}

\author{H.~M.~R{\o}nnow}
\affiliation{Laboratory for Quantum Magnetism, Ecole Polytechnique F\'ed\'erale de Lausanne (EPFL), CH-1015 Lausanne, Switzerland}

\author{B.~Normand}
\affiliation{PSI Center for Scientific Computing, Theory and Data, Paul Scherrer Institute, CH-5232 Villigen-PSI, Switzerland}

\author{M.~Mourigal}
\affiliation{School of Physics, Georgia Institute of Technology, Atlanta, GA 30332, USA}

\author{F.~Mila}
\affiliation{Institute of Physics, Ecole Polytechnique F\'ed\'erale de Lausanne (EPFL), CH-1015 Lausanne, Switzerland}

\author{A.~M.~L\"auchli}
\affiliation{Institute of Physics, Ecole Polytechnique F\'ed\'erale de Lausanne (EPFL), CH-1015 Lausanne, Switzerland}
\affiliation{PSI Center for Scientific Computing, Theory and Data, Paul Scherrer Institute, CH-5232 Villigen-PSI, Switzerland}

\author{M.~M{\aa}nsson}
\affiliation{Department of Applied Physics, KTH Royal Institute of Technology, SE-106 91 Stockholm, Sweden}

\begin{abstract}
{\bf Quantum magnets in two dimensions display strong quantum interaction effects even when magnetically ordered. Using the metal-organic framework material CuF$_2$(D$_2$O)$_2$(pyz), we investigate the field-dependent spin dynamics of the $S = 1/2$ square-lattice Heisenberg antiferromagnet by high-resolution inelastic neutron scattering to applied fields beyond one third of saturation. We discover an anomalously sharp, dispersive ``shadow mode'' residing within the two-magnon continuum, which shadows the dispersion of the transverse one-magnon branches across the Brillouin zone at an offset equal to the Larmor energy. We perform cylinder matrix-product-state (MPS) calculations that reproduce the field-induced spectrum quantitatively and apply a spectrally consistent $1/S$ spin-wave theory to deduce that the ``Larmor-shadow mode'' is a composite two-magnon resonance: a dispersing magnon at wavevector ${\bf Q}$ couples to the uniform Larmor precession at $\Gamma$, its small intrinsic linewidth indicating a non-perturbative effect of attractive magnon-magnon interactions. Another quantum-fluctuation phenomenon, the zero-field $(\pi,0)$ anomaly, is lost at increasing fields, which tighten the spectral weight into the one-magnon and Larmor-shadow modes. To our knowledge, these results constitute the first observation of a sharp quasi-bound state embedded in the continuum of a gapless two-dimensional antiferromagnet.}
\end{abstract}

\maketitle

In one spatial dimension (1D), the properties of quantum spin systems are dominated by quantum-fluctuation effects. Intensive theoretical and experimental research devoted to finding the same properties in two and higher dimensions, specifically the quantum spin-liquid (QSL) ground state and its fractional spinon and emergent gauge-field excitations,\cite{Savary2016} has focused on highly frustrated quantum magnetic models and materials. Although unfrustrated 2D and 3D quantum magnets have a finite degree of magnetic order, their properties may nevertheless be strongly renormalized by quantum fluctuations\cite{Manousakis1991} and some of their behavior may lie entirely beyond a semiclassical description. 

The $S = 1/2$ square-lattice Heisenberg antiferromagnet (SLHAF) is the keystone model of unfrustrated quantum magnetism in 2D, and has been investigated in great detail analytically, numerically and experimentally for its role in understanding the magnetic dynamics of cuprate superconductors.\cite{Lee2006} Measurements of the zero-field magnetic excitations for several representative N\'eel-ordered SLHAF compounds\cite{Keimer1992, Coldea2001, Kim2001, Ronnow2001} revealed well-defined spin waves throughout the Brillouin zone (BZ), albeit with anomalous scattering at the zone-boundary wavevectors that has been discussed in terms of extended effective spin interactions.\cite{Coldea2001, Kim2001, Sandvik2001} However, a suppression of the magnon branch around the $(\pi,0)$ point\cite{Ronnow2001, Christensen, Tsyrulin2009, Headings2010, DallaPiazza, Plumb2014} has given rise to competing theoretical proposals based on multimagnon bound or scattering states\cite{Singh1995, Sandvik2001, Zheng2005, Powalski, Powalski2018} and on magnon fractionalization.\cite{DallaPiazza,Shao2017,Yu2018,Zhang2022}

\begin{figure*}
    \includegraphics[width=\textwidth]{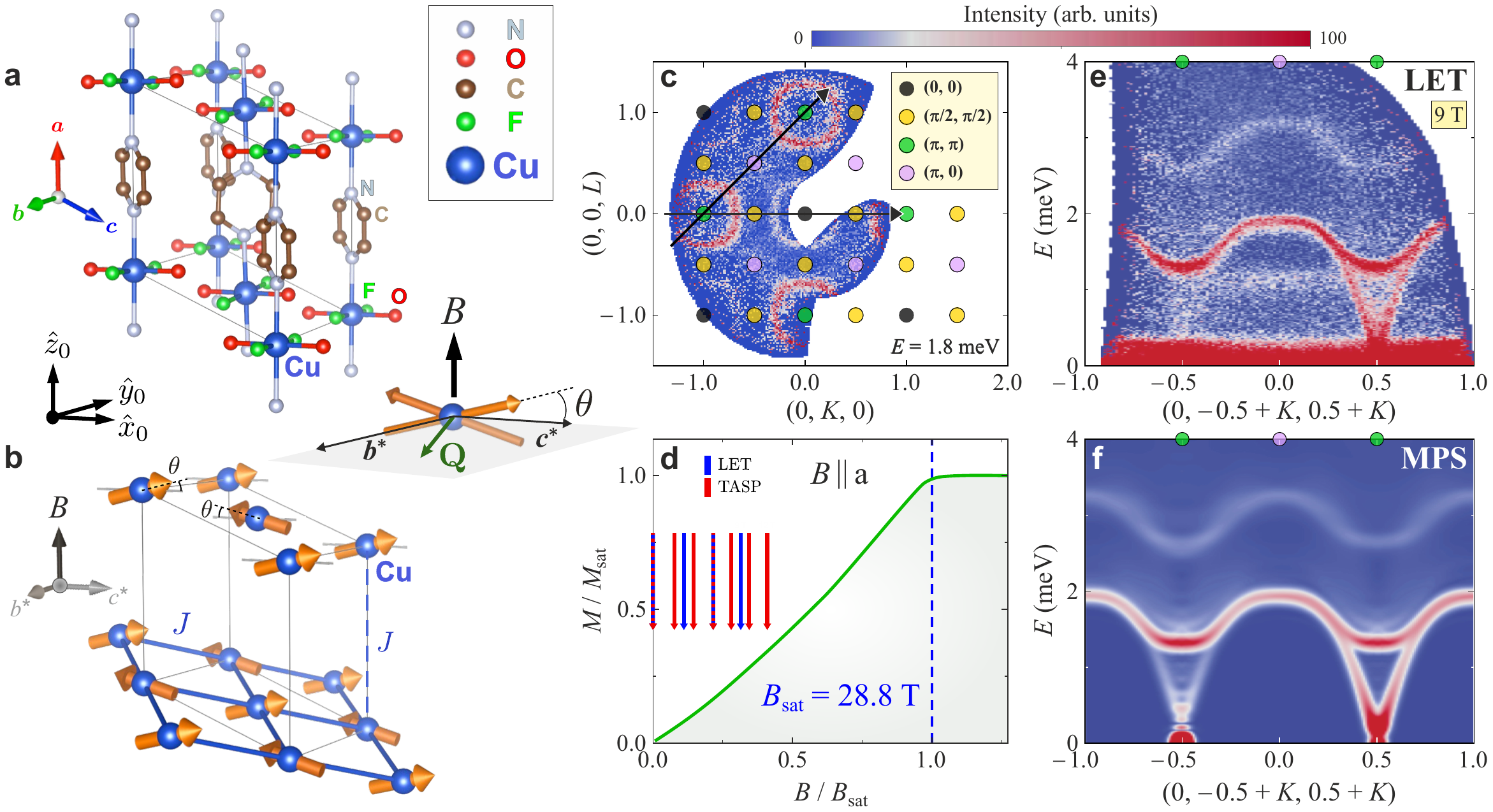}
    \caption{ {\bf CuF$_2$(D$_2$O)$_2$(pyz) and its field-induced excitation spectrum.} 
    {\bf a} Monoclinic crystal structure of CuF$_2$(D$_2$O)$_2$(pyz), highlighting its quasi-2D nature with square-lattice planes ($bc$) separated along $a$ by organic molecules. 
    {\bf b} Canted magnetic structure induced by the magnetic field, which is applied with $\hat{B} \parallel \hat{a}$. The magnetic moments are canted by angle $\theta$ towards the field direction, remaining in the $ac^\ast$ plane. The inset represents the experimental geometry, where the momentum transfer {\bf Q} lies within the $b^\ast c^\ast$ plane, and defines the Cartesian laboratory frame $(x_0,y_0,z_0)$ used in our calculations.
    {\bf c} Overview of the reciprocal-lattice plane ${\bf Q} = (0,K,L)$ illustrated by a constant-energy scan at $E = 1.8$~meV and $B = 0$~T. Coloured points mark equivalent positions and their equivalence to the reciprocal-lattice points commonly referred to as $(0,0)$, $(\pi,0)$, $(\pi,\pi)$ and $(\pi/2,\pi/2)$. The blue region is the {\bf Q} space accessible in our time-of-flight (TOF) experiment with an incident neutron energy of $E_i = 5.5$~meV. Black arrows mark the high-symmetry {\bf Q} paths shown in Fig.~\ref{fig:Comparison}. 
    {\bf d} Isothermal magnetization of CuF$_2$(H$_2$O)$_2$(pyz) at $T = 0.5$~K (data adapted from Ref.~\cite{manson_chem}). Blue arrows indicate the fields at which INS data were collected in our TOF (LET) experiment and red arrows the fields of the triple-axis (TASP) experiment. 
    {\bf e},{\bf f} Overview of spectra measured by neutron spectroscopy (LET) and computed by cylinder matrix-product states (MPS) for a field of $B = 9$~T, showing the two clear low-energy magnon branches accompanied by an emerging higher-lying mode that ``shadows'' the one-magnon branches at a {\bf Q}-independent energy offset.
    \label{fig:Overview}}
\end{figure*}

In Heisenberg models, an applied magnetic field serves as a control parameter, gradually polarizing the ordered moments and eliminating quantum fluctuations entirely in the saturated phase. The excitation spectrum of the SLHAF in a field has received relatively little attention, due in part to a lack of materials with sufficiently small energy scales for the available high-field magnets. Most theoretical studies have focused on the high-field regime, motivated by the surprising result\cite{Zhitomirsky1999} that magnons become unstable due to cubic interactions above a threshold field $B^* = 0.76~B_{\rm sat}$. Subsequent studies have extended the spin-wave-theory (SWT) description of quasiparticle decay into two-particle continua\cite{Mourigal2010,Fuhrman2012} and generalized it to valence-bond systems.\cite{Stone2005, Zhito2013, Plumb2016, Hong2017} Numerical studies using exact diagonalization (ED) found clear deviations from interacting SWT at low fields but confirmed the magnon instability at fields just below saturation.\cite{Lauchli2009} A further intriguing ED result is the emergence of a distinct excitation that moves linearly in energy above the one-magnon branch as the field is increased.\cite{Lauchli2009} That no such excitation has yet been observed in a SLHAF material, or reported in a SWT framework, poses an open challenge to theory and experiment.

The synthesis of metal-organic framework materials,\cite{Huang2024,Drichko2026} in which magnetic ions are incorporated into an organic framework to realize magnetic Hamiltonians in simple geometries, has expanded the scope for experimental studies of the SLHAF at both low and intermediate fields. In contrast to the sub-eV energy scale of the cuprates, Cu(DCOO)$_2$$\cdot$$4$D$_2$O (CFTD) \cite{Ronnow2001,Christensen,DallaPiazza} has a nearest-neighbor interaction energy of only $J = 6.3$~meV. In Cu(pyz)$_2$(ClO$_4$)$_2$ this value is 1.6~meV,\cite{Lancaster2007,Tsyrulin2009} which puts the saturation field at $B_{\rm sat} \approx 45$~T; in (5CAP)$_2$CuCl$_4$,\cite{Coomer2007} $B_{\rm sat}$ is only 3.6 T, but the small excitation bandwidth is a challenge to inelastic neutron scattering (INS). In this context, CuF$_2$(D$_2$O)$_2$(pyz) [where (pyz) denotes D$_4$-pyrazine] has $B_{\rm sat} \simeq 28.8$~T\cite{manson_chem} and hence a bandwidth ideal for high-resolution INS using cold neutrons. Early experiments on CuF$_2$(D$_2$O)$_2$(pyz) in zero field found sharp one-magnon excitations centred around the $(\pi,\pi)$ positions of the BZ, an ordered Cu$^{2+}$ moment of 0.6 $\mu_{\rm B}$, consistent with the zero-point fluctuations expected for the SLHAF, and a flat dispersion in the interplanar direction confirming good 2D character,\cite{OakRidge} making this an excellent candidate for investigating the excitation spectrum of the SLHAF at finite fields. 

In this Article, we present high-resolution INS measurements on CuF$_2$(D$_2$O)$_2$(pyz) in external magnetic fields up to $B = 0.4 B_{\rm sat}$. The field-dependence of the primary magnon modes is in excellent agreement with theoretical predictions for the SLHAF. In addition, we find a well-defined collective excitation that ``shadows'' the position of the one-magnon branch but is offset to higher energies by an amount that increases linearly with the field and places it well inside the multi-magnon continuum. We perform cylinder matrix-product-state (MPS) calculations of the spectral function that capture all the spectral features observed in experiment with quantitative accuracy, confirming both the SLHAF model and the intrinsic nature of these features. To interpret the origin of the shadow excitation, we use SWT to trace this to the combination of a field-induced Larmor mode at ${\bf Q} = 0$ and a dispersing magnon, but find that, unlike MPS, available perturbative SWT approaches cannot capture the strikingly narrow linewidth of this mode. By applying a transverse staggered field in our MPS calculations, we show how the shadow excitation is connected to a true two-magnon bound state. Given its dispersing nature, spectral origin and almost infinite lifetime, we refer to this sharp collective resonance as the Larmor-shadow mode (LSM). While bound states can appear below the continuum at certain {\bf Q} values in a gapped system, the discovery of such a coherent resonance mode in a gapless quantum magnet is unique. Our results therefore reveal yet another new and surprising effect of magnon-magnon interactions: the field-induced formation of a quasi-bound state from and within the continuum excitations.

\medskip
\noindent {\bf \large {Results}} 
\vspace{2pt}

\noindent {\bf Experiment and Analysis} 

\noindent
CuF$_2$(D$_2$O)$_2$(pyz) crystals were grown by evaporative crystallization using deuterated precursors. The sample used in these experiments was a single crystal of mass 1.1~g and dimensions 18$\times$10$\times$7~mm. Details of this process are presented in the Methods section, in Sec.~S1 of the Supplementary Information (SI)\cite{si} and also in Ref.~\cite{Lanza2014}

We performed time-of-flight (TOF) INS measurements on the LET spectrometer at the ISIS Pulsed Neutron and Muon Source. In the monoclinic structure of CuF$_2$(D$_2$O)$_2$(pyz), shown in Fig.~\ref{fig:Overview}a, the square lattice corresponds to the crystallographic $bc$ plane (Fig.~\ref{fig:Overview}b), with a superexchange interaction between nearest-neighbour Cu$^{2+}$ moments of $J = 0.934$~meV.\cite{OakRidge} Magnetic fields of $B = 0$, 3, 6 and 9~T were applied along the crystallographic $\hat{a}$ direction, which is perpendicular to the $b^*c^*$ plane in which the crystal was aligned (Figs.~\ref{fig:Overview}b,c) and has a $g$-tensor value of $g_a = 2.42$.\cite{manson_chem} Three incident neutron energies were used to balance {\bf Q}-space coverage with $E$-resolution, and most of the data we present were taken at $E_i = 5.5$ meV. To obtain a higher maximum and finer field coverage, we performed additional measurements on the triple-axis spectrometer TASP at the Paul Scherrer Institute, where fields from $B = 0$ to 12~T ($B/B_{\rm sat} \approx 0.4$) were applied in 2~T steps. From the isothermal magnetization, shown in Fig~\ref{fig:Overview}d, it is clear that these experiments probe the low-field regime up to $B/B_{\rm sat} \approx 0.4$. All measurements were performed at $T = 1.7$~K, which is below the N\'eel temperature, $T_{\rm N} = 2.6$~K.\cite{manson_chem} Further information about these experiments and about our data treatment is provided in the Methods section and in Sec.~S2 of the SI.\cite{si}

\begin{figure*}[t]
\includegraphics[width=\textwidth]{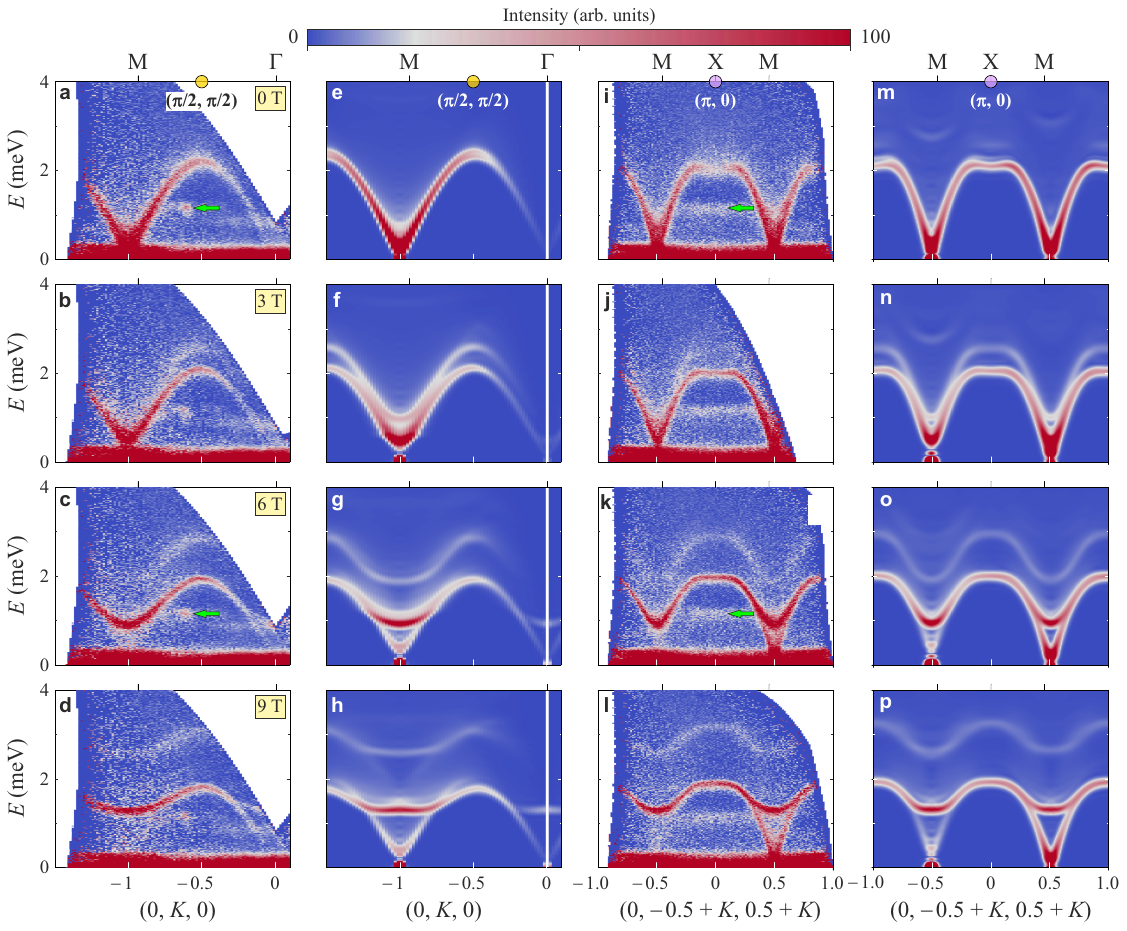}
\caption{{\bf Comparison of zero-field and field-induced INS and MPS spectra.} {\bf a}-{\bf d},{\bf i}-{\bf l} Neutron scattering intensity, $I(\mathbf{Q},E)$, measured on LET ($E_i = 5.5$~meV) for the two high-symmetry reciprocal-space directions at the four experimental magnetic fields. $(0,K,0)$ is the ${\rm \Gamma M}$ direction and includes the $(\pi/2,\pi/2)$ point. $(0,-0.5+K,0.5+K)$ is the ${\rm XMX}$ direction and includes the $(\pi,0)$ point. Data were binned using $\delta E = 0.015$~meV and $\delta Q = 0.015$~r.l.u., and were integrated in the perpendicular in-plane direction over $\Delta Q = \pm 0.1$ r.l.u.~and in the out-of-plane direction over $\Delta H = \pm 0.4$ r.l.u. We observe the progressive field-induced splitting of the one-magnon excitations into two transverse modes (corresponding to the uniform and staggered parts of the magnetization) and the development of a sharp resonance mode in the multi-magnon continuum at higher energies. The field-independent feature around 1.2 meV (green arrows) is a spurion arising from an additional reflection within the LET magnet sample environment. {\bf e}-{\bf h},{\bf m}-{\bf p} Spectral functions calculated by cylinder MPS for both directions at the same four fields, convolved with the polarization factor for direct comparison with the INS data.}
\label{fig:Comparison}
\end{figure*}

We apply cylinder matrix-product-state (MPS) methods\cite{Schollwoeck2011,Gohlke2017,Verresen2018,Xie2023} to perform a quantitative numerical analysis of the dynamical properties of CuF$_2$(D$_2$O)$_2$(pyz) for uniform external fields equivalent to $B = 0$, 3, 6 and 9~T. We do not attempt to reconsider or refine the spin Hamiltonian, for example by identifying sub-leading interaction terms, and model only the SLHAF with $J = 0.934$ meV.\cite{OakRidge} To focus on the field effects, we use a cylinder of length $L = 100$ and width $w = 6$, a tensor bond dimension of $\chi = 400$ and two different wrappings of the square lattice on the cylinder in order to access the two high-symmetry directions of the Brillouin zone (Fig.~\ref{fig:Overview}c) for a full comparison with the INS data. To establish the direction and the familiar 61\% magnitude of the ordered moment of the SLHAF at zero field in this finite system, we apply a tiny transverse staggered field, $h_x = 0.01J$, which we stress is much smaller than the energy resolution and has no effect on the calculated dynamics. A qualitative advantage of our MPS calculations is to separate the contributions to the dynamical structure factor into the three inequivalent spin channels, which we detail in full in the Methods section and in Sec.~S3 of the SI.\cite{si} 

To interpret our INS and MPS results, we consider a spectrally consistent SWT including $1/S$ corrections, as implemented in Refs.~\cite{Zhitomirsky1999,Mourigal2010,Fuhrman2012} In particular we use the SWT analysis for a systematic definition of directions and components in spin space and in real space. For this we define a laboratory frame $(\hat{x}_0, \hat{y}_0, \hat{z}_0)$, where $\hat{z}_0$ is aligned with the field, ${\hat B}$, which in turn is applied along the crystalline $a$ axis (Fig.~\ref{fig:Overview}b). Because the canted magnetic order has no component along $b^\ast$ at any field, we align ${x}_0$ with $c^*$ such that the canted spins lie in the $x_0z_0$ ($c^\ast a$ plane). For simplicity we neglect a very small tilt towards $\hat{a}$ that is present at zero field and assume that the moments are aligned in the scattering plane (along $c^\ast$) before the field induces a canting ($\theta$).The spin operators are then quantized along $\hat{z}$ in a local frame $(\hat{x},\hat{y},\hat{z})$ that rotates with the ordered moments. In this way we isolate fluctuations that are transverse to the ordered moments (${x}{x}$ and ${y}{y}$, corresponding primarily to one-magnon excitations) from those that are longitudinal (${z}{z}$, corresponding to two-magnon excitations). The components of the dynamical structure factor are then calculated in the laboratory frame, which we also used for our MPS calculations. Further details are provided in the Methods section and a complete derivation in Sec.~S4 of the SI.\cite{si} 

\noindent
{\bf Survey of reciprocal space} 

\noindent
We begin the presentation of our INS results with an overview of the reciprocal-space region accessed in the LET experiment with $E_i = 5.5$ meV. The constant-energy spectrum in Fig.~\ref{fig:Overview}c shows prototypical ``rings'' of sharp, spin-wave-like excitations originating from the ${\rm M} \equiv (\pi,\pi)$ magnetic zone centres, which at $E = 1.8$ meV start to display the square symmetry of the zone-boundary magnon dispersion of the SLHAF. With this $E_i$, repeated structural zone centers [the $\Gamma \equiv (0,0)$ points] lie on the boundary of our ${\bf Q}$ coverage and are not accessible. To characterize the field-dependence of the excitation spectra along the high-symmetry paths, in Figs.~\ref{fig:Comparison}a-d we show the scattering intensity $I(\mathbf{Q},E)$ at fields of $B = 0$, 3, 6 and $9$~T along the $(0,K,0)$ direction, which connects the $(\pi,\pi)$ point to the $(\pi/2,\pi/2)$ point, and in Figs.~\ref{fig:Comparison}i-l the intensity measured along the $(0,-0.5+K,0.5+K)$ direction, which connects the M point to the ${\rm X} \equiv (\pi,0)$ point; these two paths are indicated by the black arrows in Fig.~\ref{fig:Overview}c. 

At zero field, we observe doubly degenerate Goldstone modes emerging from the magnetic zone centres at $(0, -1, 0)$ and $(0, 0, 1)$. Between these two points (Fig.~\ref{fig:Comparison}i), we observe a clear $(\pi, 0)$ anomaly~\cite{Ronnow2001,Christensen,DallaPiazza} where the magnon branch drops slightly in energy and significantly in intensity when compared to the $(\pi/2,\pi/2)$ point (Fig.~\ref{fig:Comparison}a); we defer a more detailed comparison to Fig.~\ref{fig:TASPLET}. Although there are no sharp features at energies above the magnon branch in the zero-field measurements, diffuse intensity is visible with a rather uniform distribution to higher energies, and we will revisit this below. We remark that the 1.2 meV intensity band visible across much of {\bf Q} at all applied fields arises due to scattering from the LET magnet and is not part of the spin response; for transparency, we leave it in our data rather than attempt a subtraction, and demonstrate its nature both by changing $E_i$ on LET (Sec.~S2 of the SI\cite{si}) and by noting its complete absence in our TASP measurements. 

\begin{figure}[t]
\includegraphics[width=\columnwidth]{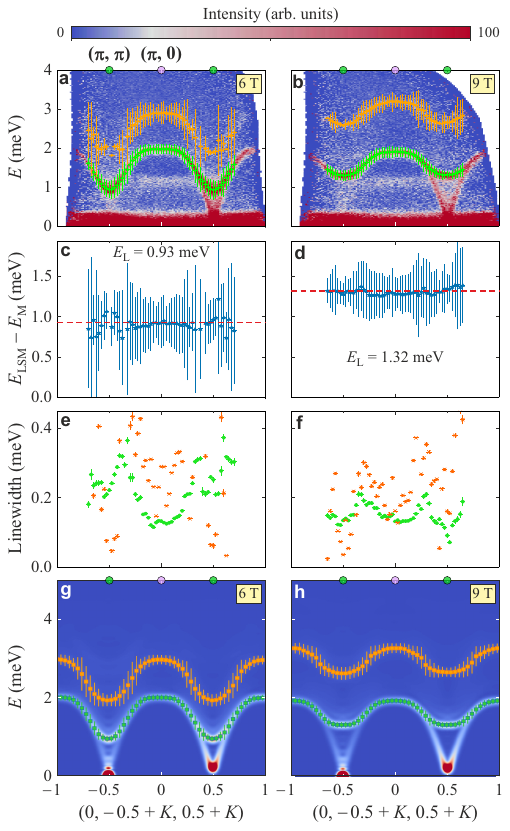}
\caption{{\bf Dispersion and linewidth of single magnons and the Larmor-shadow mode.} {\bf a},{\bf b} Dispersion relations for the gapped one-magnon branch (green) and the shadowing high-energy mode (orange) extracted from the INS intensity $I(\mathbf{Q},E)$ at $B = 6$~T (a) and $B = 9$~T (b). Symbols mark the centres of Gaussians fitted at constant ${\bf Q}$ and error bars mark the corresponding Gaussian widths. {\bf c},{\bf d} Energy difference obtained by subtracting the position of the one-magnon branch ($E_{\rm M}$) from that of the shadow mode ($E_{\rm LSM}$). The dashed purple line marks the Larmor energy, $E_{\rm L} = g \mu_{\rm B} \mu_0 B$, for the corresponding field. {\bf e},{\bf f} Full width at half-maximum height (FWHM) of excitations determined from the Gaussian fits. {\bf g},{\bf h} Dispersion and linewidth of the one-magnon and high-energy modes extracted from MPS calculations.}
\label{fig:Analysis}
\end{figure}

As a field is applied, we observe a rapid evolution of $I(\mathbf{Q},E)$, which reflects primarily the reorganization of spectral weight induced by the canting of the ordered moments and the effect of the neutron scattering polarization factor.\cite{Zhitomirsky1999,Mourigal2010,Fuhrman2012} We work in the paramagnetic Brillouin zone, where the two dynamical degrees of freedom at any wavevector {\bf Q} in the magnetic zone appear as a single one-magnon branch at momenta $\mathbf{Q}$ and $\mathbf{Q} - \mathbf{M}$. In SWT, one-magnon excitations reflect transverse moment fluctuations that can combine into staggered (out‑of‑phase) and uniform (in‑phase) modes; degenerate at zero field, these modes are mixed by the Zeeman coupling as finite fields drive the canting of the ordered moments, and appear as two distinct dispersions. The uniform component carries a net magnetization and the associated out-of-plane mode ($\parallel\!B$) is a precession appearing at the Larmor energy, $E_{\rm L} = g \mu_{\rm B} \mu_0 B$; this Larmor mode is a generic feature of Heisenberg models in an applied field and we show a measurement of it in Fig.~S7 of the SI.\cite{si} The staggered component carries minimal net magnetization and hence the in-plane ($\perp\!B$) mode changes little with field, remaining gapless.

We observe this behaviour very clearly near $(0, 0, 1)$ (Figs.~\ref{fig:Comparison}i-l), where the two previously degenerate magnons separate into distinct in-plane and out-of-plane modes. Crucially, the U(1) symmetry protects the Larmor-mode energy from renormalization due to quantum fluctuations, and our measurements of the gapped mode at $(0, -1, 0)$ (Figs.~\ref{fig:Comparison}a-d) demonstrate its linear dependence on $B$ with $g_a \simeq 2.4$. The evolution of $I(\mathbf{Q},E)$ along $(0,-0.5+K,0.5+K)$ provides an excellent confirmation of the physical picture, with the behaviour near $(0,-1,0)$ reflecting the almost complete suppression of in-plane (transverse) fluctuations by the polarization factor. This result also confirms that the moments remain perpendicular to $b^*$ at all fields (minimal leakage of intensity from the in-plane channel is caused by the out-of-plane momentum integration) and hence validates our choice for the $\hat{x}_0$ direction of the MPS staggered field.

The polarization factor along $(0,K,0)$ (Figs.~\ref{fig:Comparison}a-d) suppresses the gapless (transverse) excitations and allows us to focus on the longitudinal moment fluctuations, which in SWT correspond to two-magnon excitations. Near $(0,-1,0)$ we observe an increase in diffuse intensity, both above and below the magnon branch, which is not present near $(0,-0.5,0)$ and reflects the SWT expectation of in-plane longitudinal fluctuations at all energies down to zero at this wavevector. We remark that the intensity of the magnon branch at $(0,-0.5,0)$ [$\equiv (\pi,0)$] increases significantly with increasing field (Figs.~\ref{fig:Comparison}i-l), while the dip in its energy also flattens, clearly suppressing the anomalous behaviour at this point, and we discuss this observation more deeply below. 

\noindent
{\bf Larmor-shadow mode} 

\noindent
The most remarkable feature in the field-induced spectra is the emergence of a clearly defined mode that disperses coherently at energies above the one-magnon bands, as one observes in Figs.~\ref{fig:Comparison}b-d and \ref{fig:Comparison}j-l. Less intense than the gapped one-magnon branch, this apparent mode tracks its dispersion precisely at all ${\bf Q}$, as we show in detail in Figs.~\ref{fig:Analysis}a-b. Remarkably, the signal is sharp and clearly visible over most of the Brillouin zone at $B = 6$ and 9~T, where the mode forms a separate band within the range expected for the two-magnon continuum. Only at lower fields ($B = 3$~T in Figs.~\ref{fig:Comparison}b and \ref{fig:Comparison}j) does the identification of this mode become more challenging due to the smaller energy offset from the one-magnon branch, the presence of other weak signals and, at this field, a lower counting time in our measurements. Juxtaposed with the INS data, in Figs.~\ref{fig:Comparison}f-h and \ref{fig:Comparison}n-p we show that this mode is equally clear and sharp in our MPS spectral functions, with quantitatively identical dispersion and with vanishing intensity as {\bf Q} approaches zero.

One property of this extra mode provides the clue to its origin: its energy offset from the one-magnon branch is exactly the Larmor energy, as we show in Figs.~\ref{fig:Analysis}c-d. This indicates a composite two-magnon excitation that combines a dispersing magnon at wavevector {\bf Q} (energy $\epsilon_{\rm {\bf Q}}$) with the Larmor mode at ${\bf Q} = {\bf 0}$, creating a ``shadow'' at energy $\epsilon_{\rm {\bf Q}} + E_{\rm L}$, and hence we introduce the term LSM. The most striking property is its remarkable sharpness, and this we benchmark by direct comparison to the one-magnon excitations. Because the low-energy magnons are resolution-limited over much of the Brillouin zone (high-resolution measurements with $E_i = 2.67$ meV are shown in Sec.~S2 of the SI\cite{si}), the momentum-dependent one-magnon widths in Figs.~\ref{fig:Analysis}e-f serve as a resolution baseline that reflects the spectrometer resolution function, the field-dependent dispersion curvature and data integration effects. The width of the LSM tracks that of one-magnon excitations but is generally 30\% to 50\% wider; on average it narrows further with increasing field (Figs.~\ref{fig:Analysis}e-f), although not uniformly for all values of {\bf Q}. We therefore attribute this observation to an intrinsic linewidth and characterize the LSM as a coherent resonance dispersing within the multi-magnon continuum. Despite its two-magnon origin, this excitation is qualitatively different from the broad two-magnon continua expected in an ordered magnet,\cite{Huberman2005} and we discuss the meaning of this result below. It is also quite different from bimagnon excitations, which are zero-momentum magnon pairs common in THz spectroscopy of Heisenberg quantum magnets.\cite{Perkins1993,Betto2021}

Turning to our cylinder MPS spectra (Figs.~\ref{fig:Comparison}e-h and \ref{fig:Comparison}m-p), these are expected to be quantitatively accurate at all $B$, ${\bf Q}$ and $E$ concerning the spectral energies and qualitatively accurate concerning linewidths and intensities, which remain somewhat dependent on our calculation parameters (Sec.~S3 of the SI\cite{si}). MPS confirms that all the spectral features we observe are intrinsic to the nearest-neighbor SLHAF, and also validates our description of the canted magnetic structure, ordered moment and neutron polarization factor. As noted above, these calculations provide independent confirmation of a sharp, coherently dispersing mode appearing one Larmor energy above the one-magnon branch throughout the full ${\bf Q}$-space, and with a linewidth slightly broadened compared to one-magnon excitations (Figs.~\ref{fig:Analysis}g-h). MPS also reproduces finer experimental observations including the narrowing of the average LSM linewidth and the disappearance of the anomalous $(\pi,0)$ scattering with increasing field. 

\begin{figure*}[t]
\includegraphics[width=\textwidth]{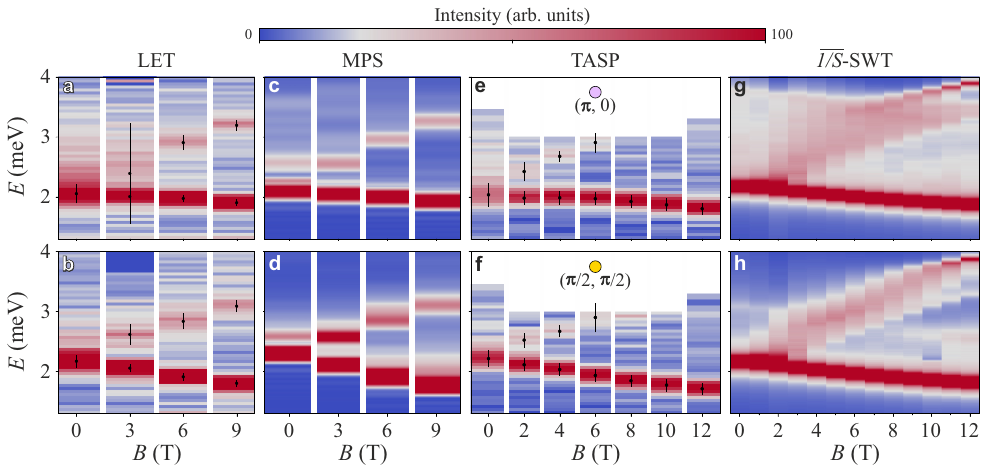}
\caption{{\bf Field-dependence of zone-boundary excitations.} Comparison of the excitation spectra at the two high-symmetry zone-boundary points $(\pi,0)$ and $(\pi/2,\pi/2)$ for all measured magnetic fields. {\bf a},{\bf b} Experimental data acquired using TOF neutron spectroscopy (LET) at $B =$ 0, 3, 6 and 9 T at ${\bf Q} = (0,-0.5,0.5) \equiv (\pi,0)$ and $(0,-0.5,1) \equiv (\pi/2,\pi/2)$. {\bf c},{\bf d} MPS calculations performed for the corresponding fields. {\bf e},{\bf f} Experimental data acquired using triple-axis neutron spectroscopy (TASP) at $B =$ 0, 2, 4, 6, 8, 10 and 12 T at ${\bf Q} = (0,0.5,-0.5) \equiv (\pi,0)$ and $(0,1,-0.5) \equiv (\pi/2,\pi/2)$. {\bf g},{\bf h} $\overline{1/S}$-SWT calculations performed for the corresponding fields.} 
\label{fig:TASPLET}
\end{figure*}
\begin{figure}[t]
\includegraphics[width=\columnwidth]{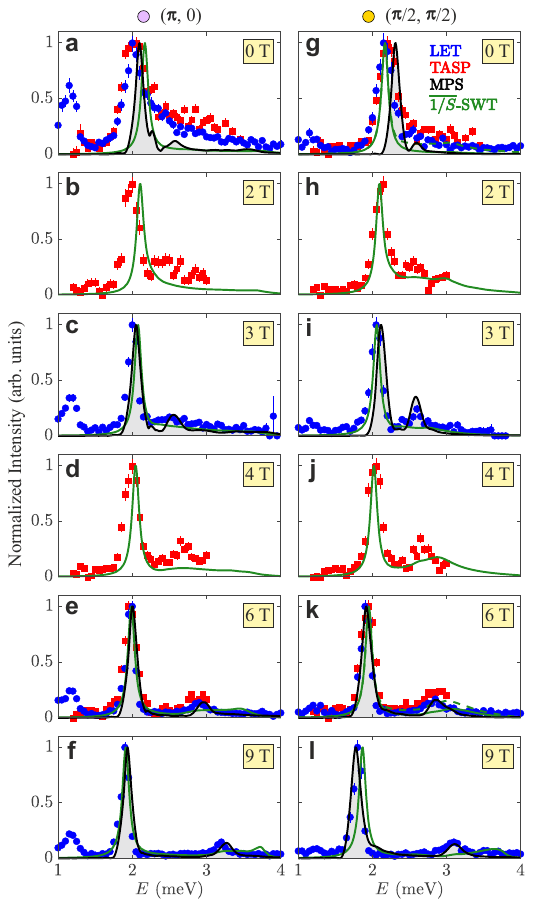}
\caption{{\bf Full experimental and calculated spectra at the high-symmetry points for all fields.} {\bf a}-{\bf f} Experimental data from ToF (LET, blue) and triple-axis (TASP, red) neutron spectroscopy compared with spectra obtained from MPS calculations (black lines with grey shadowing) and $\overline{1/S}$-SWT calculations (green lines) for the $(\pi,0)$ point. The feature at $E \approx 1.2$ meV in the LET data is an artifact of the magnet. {\bf g}-{\bf l} Corresponding spectra for the $(\pi/2,\pi/2)$ point. Because the $(\pi/2,\pi/2)$ points $(0, -0.5, 1)$ and $(0, 1, 0.5)$ have inequivalent polarization factors, the full and dashed green lines in panels g and k represent respectively the two different $\overline{1/S}$-SWT predictions for the LET and TASP experiments.}
\label{fig:LinSpin}
\end{figure}

Having established the equivalence of our INS and MPS spectra, we use the latter to separate the dynamical structure factor by spin channel, which we illustrate for the XMX path of Figs.~\ref{fig:Comparison}i-p in Sec.~S3 of the SI.\cite{si} As expected from SWT,\cite{Mourigal2010,Fuhrman2012} the gapless one-magnon mode associated with in-plane fluctuations dominates the $y_0y_0$ channel, while the gapped magnon with out-of-plane fluctuations dominates $z_0z_0$. LSMs are present in both channels, but with crossed character: the shadow in $y_0y_0$ combines a Larmor mode with a gapped single magnon, while the shadow in $z_0z_0$ has a deep energy dip at the M point, indicating the combination of a Larmor mode with a gapless magnon. We remark again on the vanishing LSM intensity as $Q \rightarrow 0$, which is another consequence of the Larmor theorem (protection by the U(1) symmetry). Other than the anomalous sharpness of the LSM, the $x_0x_0$ and $z_0z_0$ responses are consistent with SWT expectations, which include one-magnon and two-magnon contributions. However, the $y_0y_0$ channel departs from zeroth-order SWT predictions, which expect only a one-magnon contribution and miss the two-magnon modes we find in INS and MPS. We deduce that the LSM formed from the gapped one-magnon branch is always visible in our data, thus being significantly more isotropic in space than SWT would predict.  

Finally, we may also use our MPS calculations to investigate the microscopic origin and robustness of the LSM by increasing the staggered transverse field beyond the minimal value, $h_x = 0.01J$, that we applied to assure an accurate treatment of the ordered moment in CuF$_2$(D$_2$O)$_2$(pyz). As we show in Sec.~S3 of the SI,\cite{si} increasing $h_x$ gaps and flattens the one-magnon mode, which in turn compresses the two-magnon continuum. The LSM then evolves continuously from a resonance inside the two-magnon continuum into a distinct bound state of two spin-flips that separates from the continuum to lie completely below it once $h_x$ approaches $J$. From this limit, the LSM can be understood as the remnant of a true bound state that survives inside the two-magnon continuum as a well-defined resonance for reasons we elucidate below. 

\noindent
{\bf Zone-boundary excitations}

\noindent
Next we turn to a more detailed analysis of the field-dependence of the spectral function at the magnetic zone-boundary points $(\pi,0)$ and $(\pi/2,\pi/2)$. Figure \ref{fig:TASPLET} shows the comparison between line cuts through our LET data at constant $\mathbf{Q}$ and the corresponding spectra collected for a finer field grid using TASP. The LSM is present with the same properties in both datasets. Given the fundamentally different nature of spurious features in these experiments, and the fact that MPS provides a near-unbiased numerical calculation of the spectral function, an extrinsic origin for the mode is completely excluded.

Quantitatively, Fig.~\ref{fig:TASPLET} shows the downward energy renormalization and sharpening of the one-magnon branch as the field is increased. This occurs non-linearly and with increasing downward curvature at $(\pi,0)$ (Figs.~\ref{fig:TASPLET}a,c,e), while the decrease is stronger but more linear at $(\pi/2,\pi/2)$ (Figs.~\ref{fig:TASPLET}b,d,f). For both points, the emergence of the LSM is not simply a sharp mode splitting off, but rather a gradual field-induced process in which one broad spectral feature around the zero-field intensity peak tightens up into two clearly defined modes (the one-magnon and Larmor-shadow branches). This process is more clearly visible at $(\pi,0)$, where the two modes are not yet fully separated at $B = 3$~T (Fig.~\ref{fig:TASPLET}a), but is also evident at $(\pi/2,\pi/2)$ at $B = 2$~T (Fig.~\ref{fig:TASPLET}g)

This evolution is also visible in the constant-$\mathbf{Q}$ spectra presented in Fig.~\ref{fig:LinSpin}, where at $B > 3$~T, when the two modes are well separated, the MPS spectral function is in quantitative agreement with the experimental data (Figs.~\ref{fig:LinSpin}e,f,k,l). This is also true at $B = 3$~T for $(\pi/2,\pi/2)$ (Fig.~\ref{fig:LinSpin}i), but for $(\pi,0)$ at 3~T (Fig.~\ref{fig:LinSpin}c) and at zero field (Figs.~\ref{fig:LinSpin}a,g) our MPS calculations no longer capture the positions and linewidths of the experimental features. This result indicates that the origin of the anomalous zone-boundary scattering in the SLHAF\cite{Ronnow2001, Christensen, Tsyrulin2009, Headings2010, DallaPiazza, Plumb2014} lies in long-ranged entanglement, which cannot be captured adequately on a $w = 6$ or even a $w = 8$ cylinder.\cite{Verresen2018} Concerning the $(\pi,0)$ anomaly, our observation that increasing field restores a somewhat ``conventional'' response, in the form of separate one- and two-magnon sectors beyond $B \approx 0.1 B_{\rm sat}$, indicates either a field-induced restoration of the spectral-weight hierarchy of multimagnon resonances\cite{Singh1995, Sandvik2001, Zheng2005, Powalski, Powalski2018} or a field-induced reconfinement of spinons resulting from zero-field magnon fractionalization.\cite{DallaPiazza,Shao2017,Yu2018,Zhang2022}

\noindent
{\bf Spin-wave theory} 

\noindent
Our combined INS and MPS results validate the description of CuF$_2$(D$_2$O)$_2$(pyz) using the SLHAF at fields up to $B = 0.4 B_{\rm sat}$. These results make clear that the LSM is a well-defined feature of the excitation spectrum of the SLHAF. Given our experimental deduction that it is a composite excitation involving a dispersing magnon and a Larmor mode, it is important to compare our data with SWT, which provides an interpretable framework for the construction of two-magnon states,\cite{Huberman2005} and hence for understanding their contributions to the field-induced neutron scattering response.\cite{Zhitomirsky1999,Mourigal2010,Fuhrman2012} Most surprisingly, the LSM has not previously been remarked upon within SWT, the only hint of its possible existence coming from numerical methods.\cite{Lauchli2009} 

Given that one-magnon energies in an applied field are renormalized by cubic vertices,\cite{Mourigal2010,Fuhrman2012} and our need to describe the energies of one- and two-magnon states equally, we perform SWT at order $1/S$ with spectral consistency,~\cite{Zhitomirsky1999,Chernyshev2009} which we denote by $\overline{1/S}$-SWT. In this approach, detailed in full in Sec.~S4 of the SI,\cite{si} the two-magnon spectrum is computed using the dispersion of the renormalized one-magnon states, calculated on-shell for consistency. In addition, we compare the results for the spectral function of one-magnon states, calculated both on- and off-shell, the latter capturing potential transverse sidebands, where spectral weight from the one-magnon pole is transferred to the higher-energy continuum. The results of $\overline{1/S}$-SWT with magnon sidebands are depicted in Figs.~\ref{fig:TASPLET}d and \ref{fig:TASPLET}h, and are represented by the green lines in Fig.~\ref{fig:LinSpin} for the two high-symmetry zone-boundary points $(\pi,0)$ and $(\pi/2,\pi/2)$. Where linear SWT predicts field-independent one-magnon energies at both points, $\overline{1/S}$-SWT reproduces the observed decrease in energy with increasing field. However, $\overline{1/S}$-SWT continues to predict a broad and continuous two-magnon response formed from these renormalized single magnons; although the SWT continuum does have its strongest intensity at the energy of the LSM (a logarithmic van Hove singularity that results from combining the local minimum at the Larmor mode with a dispersing magnon, as we discuss in Sec.~4 of the SI\cite{si}), there is no tendency towards the formation of a sharp resonance. $\overline{1/S}$-SWT does, however, account for the near-isotropy of the LSM in spin space found by MPS, in that a significant transfer of spectral weight to higher energies occurs within the transverse channel (a result we demonstrate in Fig.~S12 of the SI\cite{si}).

It is crucial to note that $\overline{1/S}$-SWT does not compute corrections beyond the dispersion of individual magnons. Our calculation therefore benchmarks a non-interacting multiparticle continuum, and signals clearly that the origin of the LSM lies in magnon-magnon interactions. Because the magnon density is small at low energies, the quartic vertices responsible for two‑magnon scattering usually play only a minor role at zero field, but at finite fields the magnon spectral density becomes concentrated at the Larmor energy. We conjecture that the U(1) symmetry protecting the Larmor mode from energetic renormalization eliminates level repulsion with the partner magnon, generating an effective attraction that converts the underlying van Hove singularity into a pole inside the two‑magnon continuum. In this scenario, the sharp resonance reflects a non‑perturbative enhancement of ladder type in the two‑magnon channel. The slight field-induced sharpening of the LSM also implies an increase in the attractive magnon-magnon interactions. To the extent that the $(\pi,0)$ anomaly can be considered as a superposition of multimagnon resonances, the field-driven strengthening of the two-magnon attraction erases the anomaly by concentrating the spectral weight into the one- and two-magnon states observed in Fig.~\ref{fig:TASPLET}.

Despite all of this insight, the observation of a quasi-bound state in the multiparticle continuum of a gapless quantum spin system is remarkable. Although sharp (resolution-limited) multimagnon bound states have been observed detaching from the continuum in many gapped quantum magnets, and their properties have been computed with quantitative accuracy,\cite{Ward2017,Legros2021} finding such a coherent resonance mode in a gapless system is unique. Physically, bound states in a gapped system can appear below the continuum at certain {\bf Q} values where scattering is forbidden by the kinematic conditions. In contrast, a gapless system possesses no such {\bf Q} regions and hence a quasi-bound state can only form within the continuum by bootstrapping the continuum spectral weight. 

\medskip
\noindent {\bf \large {Discussion}} 
\vspace{2pt}

We have discovered and characterized a new type of collective excitation in the SLHAF, a well-defined field-induced dispersive mode that we identify as a quasi-bound state of a single magnon and a Larmor mode. Its existence implies the presence of strong magnon-magnon interactions that are non-perturbative in a SWT framework and hence have been missed by these techniques. Systematic methods for the quantitative calculation of multi-magnon response functions\cite{Powalski,Powalski2018} have yet to be applied to the phases of canted AF order encountered at finite applied fields. Because our study was performed at fields $B \leq 0.4 B_{\rm sat}$, we do not see any of the magnon decay proposed in the SLHAF,\cite{Zhitomirsky1999} which has been analysed by SWT\cite{Mourigal2010,Fuhrman2012} and ED methods\cite{Lauchli2009} and measured in a spatially anisotropic metal-organic system.\cite{Hong2017} To the contrary, in this field range we observe a field-induced increase of peak intensities in the one-magnon branch.

The discussion of collective many-body phenomena in the SLHAF is strongly influenced by the proposal of magnon fractionalization as an explanation for the anomalous spectrum at the $(\pi,0)$ point in zero field.\cite{DallaPiazza,Shao2017,Yu2018,Zhang2022} At finite fields, and to analyse the LSM, we find no need of a spinonic description for our INS and MPS data. In fact we observe a systematic field-induced narrowing of both the one-magnon branch and the LSM at low fields. Thus, if the $(\pi,0)$ point were to host a true deconfinement at $B = 0$, then our results indicate that the spinons become reconfined at finite $B$, just as they are at finite wavevector shifts away from $(\pi,0)$.\cite{DallaPiazza,Shaik2025}

It is instructive to compare our results for the SLHAF with the other paradigm model for unfrustrated 2D quantum magnetism, the $S = 1/2$ Heisenberg antiferromagnet on the honeycomb lattice (HLHAF), which has the even smaller coordination number $z = 3$ instead of 4. Analytical studies of the excitation spectrum in finite fields have been performed only at the semiclassical level using a nonlinear SWT,\cite{Maksimov2016} while numerical approaches have focused largely on searching for magnon fractionalization at zero field, with and without frustrating interactions.\cite{Ferrari2020,Gu2022} On the experimental side, the lanthanide trihalides YbBr$_3$\cite{Wessler2020} and YbCl$_3$\cite{Sala2021} have recently been shown to provide excellent realizations of the pseudospin-1/2 HLHAF with saturation fields of approximately 9 T. The zero-field spectra of both materials show extensive continuum scattering at certain zone-boundary wavevectors.\cite{Wessler2020,Sala2021} INS measurements at 2 and 4 T in YbCl$_3$\cite{Sala2023} display an analogue of the LSM throughout much of {\bf Q}-space, but it remains a rather broad and continuum-like feature that never achieves the sharpness we observe in the SLHAF. Very recently, a comprehensive study of YbBr$_3$ combining INS at fields up to saturation with cylinder MPS calculations of the full spectrum\cite{hlsm2025} has verified the HLHAF Hamiltonian, discovered a roton-like collective mode, observed near-complete magnon decay\cite{Zhito2013,Maksimov2016} at high fields and quantified the appearance of a broadened ``magnon shadow'' at low fields. These results demonstrate that Larmor-shadow physics may be a universal feature of unfrustrated quantum spin systems, but leaves open the reasons for the very strong magnon attraction apparently at work in the SLHAF. 

We note here that the broadened shadow feature in the HLHAF tends to disappear at fields $B > 0.5 B_{\rm sat}$, and we also expect this be the case for the LSM in the SLHAF. Although this field regime has yet to be probed by experimental or numerical methods, $\overline{1/S}$-SWT calculations show that the one-magnon dispersion branches change their form quite dramatically beyond $B = 0.5 B_{\rm sat}$, as we show in Fig.~S17 of the SI.\cite{si} Under these circumstances, it is highly likely that the kinematic conditions giving rise to the very strong magnon-magnon attraction are lost and the LSM ceases to exist. 

Our results raise multiple questions about the nature of the LSM, which lead us to anticipate that they will spark a new round of theoretical and experimental analysis of the generalized SLHAF. Although it is not easy to find new SLHAF materials with small saturation fields, the LSM can certainly be studied at $B = 0.2 B_{\rm sat}$, and hence in known systems with saturation fields up to approximately 60~T. It is definitely valuable to study the evolution of the LSM as the magnon kinematics are changed by a second-neighbour coupling, $J_2$,\cite{Gong2014,Morita2015,Melzi2001,Ishikawa2017} and as they are changed by breaking the symmetry of the square lattice down to a rectangular lattice, meaning with $J_a \ne J_b$;\cite{Miyazaki1995,Shaik2025} one route to the latter type of system would be the application of a uniaxial pressure.\cite{Kim2018,Guguchia2020} Under a hydrostatic pressure, CuF$_2$(D$_2$O)$_2$(pyz) shows a significant reduction,\cite{Skoulatos2017} and the companion material (CuF$_2$(D$_2$O)$_2$)$_2$(pyz) a very dramatic reduction, of its in-plane interaction, $J$.\cite{Wehinger2018} In both materials this drop is followed by a first-order transition that switches the dimensionality to a system of quasi-1D spin chains running perpendicular to the square-lattice planes,\cite{Skoulatos2017,Wehinger2018} and hence a switch of its excitations from magnons to spinons.\cite{Skoulatos2017} Pressure control of superexchange interactions has already been effected in cuprate systems,\cite{Wang2022,Thomarat2024} and in (CuF$_2$(D$_2$O)$_2$)$_2$(pyz) could in principle be used to reduce $J$ to the extent that the magnon-decay regime becomes accessible. 

In conclusion, we have utilized INS measurements in combination with MPS and spectrally consistent SWT calculations to investigate the field-induced spin excitations of the SLHAF material CuF$_2$(D$_2$O)$_2$(pyz). We show the emergence of a sharp collective mode that shadows the one-magnon branch at an upward energy offset given exactly by the Larmor frequency. This novel Larmor-shadow mode (LSM) went previously unnoticed in SLHAF studies and measurements, and we show that its remarkably narrow width is not captured appropriately by available perturbative SWT methods, but is captured by cylinder MPS calculations, which include all quantum fluctuation effects in a way that is not biased by an Ansatz. Our results constitute the first discovery of a quasi-bound state in the continuum of a gapless quantum magnet and hence represent qualitative progress in understanding the many-body response of gapless spin systems. 

\medskip
\noindent
\noindent {\bf \large {Methods}} 
\vspace{2pt}

\vspace{4pt}
\noindent
{\bf {Materials synthesis.}} Large and high-quality crystals of CuF$_2$(D$_2$O)$_2$(pyz) were grown by an evaporative growth technique, using the method described in Ref.~\cite{Lanza2014} with deuterated counterparts of the chemicals involved. Here, 12.91 g of copper(II) chloride were dissolved in 192~ml of water with an additional 32.62~g of silver nitrate solution in 96 ml of water. This mixture was cooled to 5$^\circ$C after the precipitate was filtered off. Then 7.11~g of ammonium fluoride and 7.69~g of pyrazine in 96~ml of water were added at this temperature, with the final solution being collected and divided into 4-6 beakers. This was then evaporated slowly at 5$^\circ$C. One solitary piece of single-crystalline sample with $m = 1.1$~g and dimensions 18$\times$10$\times$7~mm was used for all INS experiments.

\vspace{4pt}
\noindent
{\bf {Neutron Spectroscopy.}} The TOF INS experiment was performed on the LET spectrometer at the ISIS spallation neutron source, located at the Rutherford Appleton Laboratory, UK. LET is a multi-chopper spectrometer designed for high energy resolution $(\geq 0.8\% \, \delta E / E_i)$ in the cold-neutron regime.\cite{LET_Description} The five-chopper system allows INS data corresponding to multiple incident neutron energies to be collected simultaneously, and here we used the incident energies $E_{\rm i} = 2.67$, 5.5 and 13.8~meV. The magnet was a split-bore solenoid,\cite{LET_Description} which could reach a field of $B = 9$~T. All data were acquired at the base temperature $T = 1.7$~K. The single-crystalline sample was placed onto a thin, perforated (to minimize holder material), ultra-pure Al holder using thin Al wire (shown in Fig.~S1) and aligned within the $(0~K~L)$ scattering plane. Collecting one dataset required multiple sample rotation angles, which were spaced equally at steps of $\phi = 2^\circ$. Measurements at all fields other than $B = 3$~T, which covered fewer angular positions ($\phi = 4^\circ$), captured a similar volume of ({\bf Q},$E$)-space (shown in Figs.~\ref{fig:Overview}c and \ref{fig:Comparison}).

The TOF data were treated and analysed using the Horace software.\cite{Ewings2016} All figures in the manuscript and SI showing $I$({\bf Q},$E$) were prepared in energy steps of size $\Delta E = 0.015$~meV and momentum steps of size $\Delta Q = 0.015$~r.l.u. The data were integrated over windows of total width 0.8 r.l.u.~in the direction out of the scattering plane ($a^*$) and 0.2 r.l.u.~in the perpendicular in-plane direction. The width of the energy-integration window for the constant-$E$ slice shown in Fig.~1c is 0.2~meV and in Fig.~S2 it is 0.3~meV (here the out-of-plane momentum-integration width is 0.3 r.l.u.~in all three panels). 

Additional INS data were acquired using the cold-neutron triple-axis spectrometer TASP and the thermal-neutron triple-axis spectrometer EIGER at the Paul Scherrer Institute. The sample alignment (scattering plane $(0~K~L)$ and sample mount) was kept exactly the same as for the LET measurements, with the field applied along the crystallographic $a$ axis. The fields used were $B = 0$, 2, 4, 6, 8, and 12~T, and the temperature was maintained at $T = 1.7$~K using the MA15 cryomagnet. The spectrometer was set up to use final neutron wavevector $k_f = 1.3$~\AA$^{-1}$, giving an instrumental energy resolution $\Delta E \approx 85~\mu$eV. Constant-$\mathbf{Q}$ scans were collected at the points ${\bf Q} = (0,0.5,-0.5)$ ($\equiv (\pi,0)$) and ${\bf Q} = (0,1,-0.5)$ ($\equiv (\pi/2,\pi/2)$). All data from TASP were analysed using Python. 

\vspace{4pt}
\noindent
{\bf {MPS calculations.}} The method of matrix-product states (MPS) uses a tensor assigned to every site of a 1D quantum system to build a variational Ansatz that can be used to represent the wavefunction with an accuracy controlled by the tensor bond dimension, $\chi$.\cite{Schollwoeck2011} The approach of representing a 2D quantum system by wrapping the lattice onto a cylinder\cite{Stoudenmire2012} was subsequently combined with time-evolution techniques to provide a means for the quantitative computation of near-unbiased spectral functions\cite{Gohlke2017,Verresen2018,Xie2023} by due benchmarking of the cylinder circumference, $w$, and the cylinder length, $L$, which controls the maximum evolution time before undue entanglement growth. 

The spin physics of CuF$_2$(D$_2$O)$_2$(pyz) is described by the SLHAF with a nearest-neighbor Heisenberg interaction of $J = 0.934$ meV. MPS calculations were performed at fields equivalent to 0, 3, 6 and 9 T, using a cylinder of length $L = 100$ to allow sufficiently long evolution times, a cylinder width $w = 6$ and a tensor bond dimension of $\chi$ = 400. The spectral function was calculated in a fixed laboratory frame, $(\hat{x}_0,\hat{y}_0,\hat{z}_0)$, with $\hat{z_0}$ aligned along $\hat{B}$ and the spins aligned in the $x_0z_0$ plane (Fig.~\ref{fig:Overview}b). The three separate spin channels are then one out-of-plane channel ($z_0z_0$ fluctuations, $\parallel\!B$) and two in-plane channels ($x_0x_0$ and $y_0y_0$ fluctuations, $\perp\!B$). The LSM appears as a distinct feature in the transverse components ($\mathcal{S}^{x_0x_0}$ and $\mathcal{S}^{y_0y_0}$), whereas in the longitudinal component ($\mathcal{S}^{z_0z_0}$) it is more diffuse. The small, staggered field applied along the $\hat{x}_0$ direction causes the one-magnon branch to lose intensity in $\mathcal{S}^{x_0x_0}$ compared to $\mathcal{S}^{y_0y_0}$, but the intensity of the LSM remains unchanged in both. Multiplying by the appropriate polarization factors causes the intensity of the gapless (Goldstone) magnon branch to vanish in the $y_0y_0$ spin channel at $(0,-1,0)$, explaining the asymmetric distribution of intensity around $(0,-1,0)$ and $(0,1,0)$ in our INS results (Fig.~\ref{fig:Comparison}). The polarization factor for the $\mathcal{S}^{z_0z_0}$ component is finite for all \textbf{Q}, meaning that the gapped magnon branch always appears in the spectra. 

\vspace{4pt}
\noindent
{\bf {Spin-wave theory.}} We compute the dynamical spin structure factor for the quasi‑2D SLHAF with nearest‑neighbour in‑plane exchange $J$ and weak interplane exchange $J^\prime = 0.01J$, in a field applied along the laboratory axis $\hat z_0 \parallel {\hat a}$. The ordered state is treated as a canted antiferromagnet with propagation vector ${\bf Q}_m = (\pi,\pi,\pi)$ and canting angle $\theta$ (Figs.~\ref{fig:Overview}a,b). We denote contributions to the structure factor from in-plane spin fluctuations by $\mathcal{S}^{\alpha\alpha}_{\perp B}$ and from out-of-plane ones by $\mathcal{S}^{\alpha\alpha}_{\parallel B}$, where $\alpha = (x_0,y_0,z_0)$. For transverse modes, these contributions can be calculated in the local frame by shifting the spectral response from wavevector ${\bf Q}$ to $\mathbf{Q} - \mathbf{Q}_m$, and by $+ \mathbf{Q}_m$ for the longitudinal modes.\cite{Mourigal2010} Because the ordered moments of CuF$_2$(D$_2$O)$_2$(pyz) have no components along $b^\ast$ in zero field (the moments are tilted by $35^\circ$ from $\hat{c}^\ast$ into the $a^\ast c^\ast$ plane),\cite{OakRidge} we assume that this property persists for all field values and align the moments along $\hat{c}^\ast$ at $B = 0^+$ before canting them by an angle $\theta$ towards $\hat{a}$ for $B > 0$. Taking into account the polarization factor for magnetic scattering, which filters out fluctuations along $\hat{{\bf Q}}$, we express the spectral response of CuF$_2$(D$_2$O)$_2$(pyz) as the linear combination of the three terms \\[4px]
\begin{tabular}{lll}
~~$z_0z_0$ channel:~~ &    & $\big[ \cos^{2}\theta\,\mathcal{S}_{\parallel B}^{xx} + \sin^{2}\theta\,\mathcal{S}_{\parallel B}^{zz} \big]$, \\[2px]
~~$y_0y_0$ channel:~~ & $(1 \! - \! \hat{Q}_{b^\ast}^2)$ $\!\!\!\!$ & $\!\!$ $\big[ \mathcal{S}_{\perp B}^{yy} \big]$, \\[2px]
~~$x_0x_0$ channel:~~ & $(1 \! - \! \hat{Q}_{c^\ast}^2)$ $\!\!\!\!$ & $\!\!$ $\big[ \cos^{2}\theta\,\mathcal{S}_{\perp B}^{zz} + \sin^{2}\theta\,\mathcal{S}_{\perp B}^{xx} \big]$,
\end{tabular}
where $\hat{Q}_{b^\ast}$ and $\hat{Q}_{c^\ast}$ are the normalized projections of ${\bf Q}$ along the unit vectors $\hat{b}^\ast$ and $\hat{c}^\ast$. 

To obtain the one‑magnon energies, we use the on‑shell $1/S$ scheme that includes Hartree-Fock (quartic) and canting‑angle corrections as well as the one‑loop self‑energies from cubic vertices, i.e.~$\bar\varepsilon_{\mathbf{k}} = \varepsilon_{\mathbf{k}} + \delta\varepsilon^{\mathrm{HF}}_{\mathbf{k}} + \delta\varepsilon^{\theta}_{\mathbf{k}} + \Sigma_{31}(\mathbf{k},\varepsilon_{\mathbf{k}}) + \Sigma_{32}(\mathbf{k},\varepsilon_{\mathbf{k}})$. This expression is appropriate in the present low‑field regime, where the magnons remain stable; we also calculate off‑shell solutions and obtain consistent results. The transverse components of the structure factor are constructed from the single‑pole spectral function $\bar A(\mathbf{k},\omega)$ using the standard Bogoliubov coefficients and moment‑reduction factors, while the longitudinal component is evaluated as the two‑magnon density of states built from the renormalized one‑magnon dispersion, ensuring spectral consistency between transverse and longitudinal channels. We also analyse the two‑magnon density of states for van Hove singularities and identify the logarithmic enhancement using standard approaches.

\bibliographystyle{naturemag}
\bibliography{bibliography}

\vspace{12pt}
\noindent
{\bf{Acknowledgements}}
\newline
\noindent
This research was funded by the Swedish Foundation for Strategic Research (SSF) within the Swedish National Graduate School in Neutron Scattering (SwedNess), as well as by the Swedish Research Council VR (Dnr.~2021-06157, Dnr.~2022-03936, Dnr.~2025-07622 and Dnr.~2025-08127) and the Carl Tryggers Foundation for Scientific Research (CTS-22:2374). We acknowledge the financial support of the Swiss National Science Foundation under Project Nos.~200020\_150257 and 200020\_172659. The work of N.B.C. was supported by the Danish National Committee for Research Infrastructure (NUFI) through the ``ESS-Lighthouse'' Q-MAT and the instrument centre DanScatt. The work of M.Mo.~was supported by the US Department of Energy, Office of Science, Basic Energy Sciences, Materials Sciences and Engineering Division under award DE-SC-0018660. We are grateful to the ISIS Neutron and Muon Source (under Proposal No.~RB1510477) and the Swiss Spallation Neutron Source (under Proposal Nos.~20131663 and 20140654) for providing beam time for the experiments in this study, as well as for their valuable technical and scientific support. MPS calculations were performed at the Swiss Federal Technology Institute of Lausanne (EPFL) on the Jed cluster. 

\bigskip
\noindent
{\bf {Data availability}}
\newline
\noindent
Raw INS data taken at ISIS are available at \url{https://doi.org/10.5286/ISIS.E.RB1510477}. All processed INS data from LET and TASP, together with all MPS and SWT data shown in the figures of the manuscript and SI, will be made available at the time of publication on \url{https://hdl.handle.net/1853/72971}.

\bigskip
\noindent
{\bf {Code availability}}
\newline
\noindent
INS data taken at ISIS were processed using MANTID\cite{Arnold2014} and HORACE.\cite{Ewings2016} The codes for our MPS calculations are available upon request from the corresponding author.

\bigskip
\noindent
{\bf {Author contributions}}
\newline
\noindent
The experimental project was initiated by M.Mo., Ch.R.~and H.M.R. The theoretical analysis was initiated by A.M.L.~and F.M. Single crystals were grown by C.F.~and K.W.K. INS experiments were prepared and performed by M.M{\aa}., M.S., N.B.C.~and S.W., with the support of U.S. as local contact at PSI and D.V.~as local contact at ISIS. MPS calculations were performed by M.N.~with support from F.M., A.M.L.~and B.N. A complete analysis of all INS and MPS data was performed by F.E.~and A.E. SWT calculations were performed by F.E.~with support from M.Mo. The manuscript was written by F.E., M.N., M.Mo, F.M.~and B.N.~with input from all authors. Questions concerning the experimental parts of this work should be addressed to M. M{\aa}nsson, while inquiries concerning the theory and modelling should be made to F. Mila (MPS) and M. Mourigal (SWT).

\bigskip
\noindent
{\bf Competing financial interests:} the authors declare no competing financial interests.

\newpage

\setcounter{figure}{0}
\renewcommand{\thefigure}{S\arabic{figure}}
\setcounter{section}{0}
\renewcommand{\thesection}{S\arabic{section}}
\setcounter{equation}{0}
\renewcommand{\theequation}{S\arabic{equation}}
\setcounter{table}{0}
\renewcommand{\thetable}{S\arabic{table}}
\def\SCBO{SrCu$_2$(BO$_3$)$_2$ }

\onecolumngrid

\noindent
{\large {\bf {Supplementary Information}}}

\vskip4mm

\noindent
{\large {to accompany the article}}

\vskip4mm

\noindent
{\large {\bf Field-induced quasi-bound state within the two-magnon continuum of a\\square-lattice Heisenberg antiferromagnet}}

\vskip4mm

\noindent
F. Elson, M. Nayak, A. A. Eberharter, M. Skoulatos, S. Ward, N. B. Christensen, D. Voneshen, C. Fiolka, \\ K. W. Kr\"amer, Ch. R\"uegg, H. M. R{\o}nnow, B. Normand, M. Mourigal, F. Mila, A. M. L\"auchli and M. M{\aa}nsson

\vskip6mm

\twocolumngrid

\section{Crystal and Experimental Geometry}
\label{ss1}

Large single crystals of CuF$_2$(D$_2$O)$_2$(pyz) were grown as detailed in the Methods section. Powder X-ray diffraction on crushed crystals was used to determine their phase purity and to perform an initial Rietveld structural refinement, as described in Ref.~\onlinecite{Lanza2014} A further refinement of the crystal structure was performed on elastic scattering cuts from our time-of-flight data; this verified the monoclinic P2$_1$/c space group with lattice parameters $a = 7.56$ Å, $b = 7.44$ Å, $c = 6.82$ Å, $\alpha = \gamma = 90.00^\circ$ and $\beta = 113.66^\circ$. 

\begin{figure}
    \includegraphics[width=0.85\columnwidth]{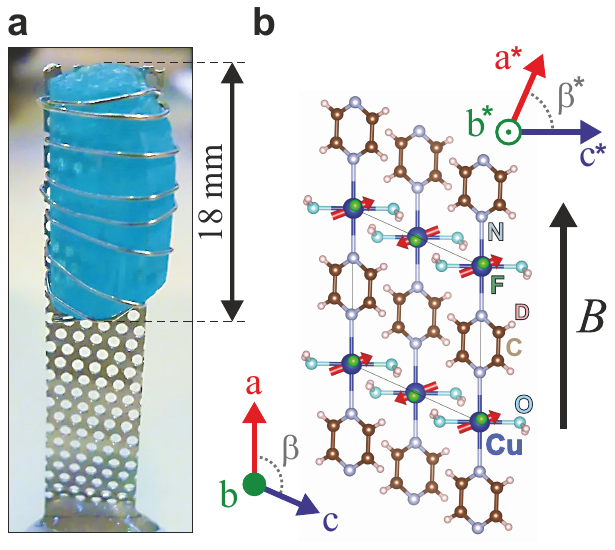}
    \caption{{\bf Single-crystal mount and crystal structure of CuF$_2$(D$_2$O)$_2$(pyz).} {\bf a} Photograph of the sample used in our scattering experiments. {\bf b} Representation of the stacking of square-lattice planes in the crystal structure, viewed along the $b$ axis and with the zero-field magnetic structure represented by the red arrows.} 
    \label{fig:CrystalStructure}
\end{figure}

Figure \ref{fig:CrystalStructure}a shows a photograph of our single-crystal sample aligned on an aluminium sample holder with the $b^\ast c^\ast$ plane [$(0~K~L)$ plane] horizontal and the magnetic field along the $a$ axis ($[100]$ axis). Figure \ref{fig:CrystalStructure}b shows a real-space representation of the crystal viewed along the $b$ axis ($[010]$ axis). The square lattice of interacting Cu$^{2+}$ ions is the $bc$ plane of CuF$_2$(D$_2$O)$_2$(pyz), with the weak interplane interactions along the $a$ axis. The zero-field magnetic structure of the material, which includes a tilting of the spins out of the $bc$ plane and towards the $a$ axis, is shown by the red arrows.

\begin{figure*}
    \includegraphics[width=0.88\textwidth]{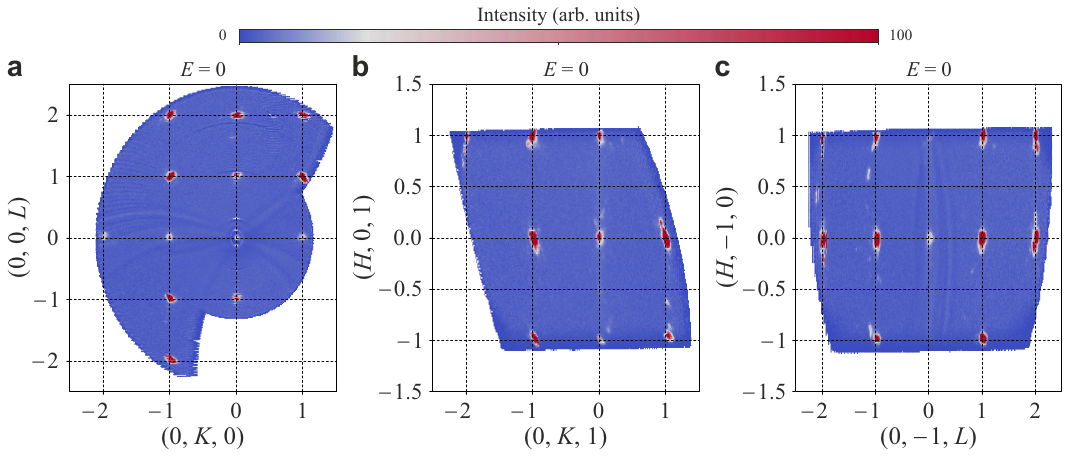}
    \caption{{\bf Quality of single-crystal alignment on LET.} Elastic neutron scattering intensity from our CuF$_2$(D$_2$O)$_2$(pyz) crystal for incident energy $E_i = 13.85$~meV, sliced in the {\bf a-c} $(0,K,L)$, $(H,K,1)$ and $(H,-1,L)$ planes, respectively, covering several repeated Brillouin zones. } 
    \label{fig:ElasticAlignment}
\end{figure*}

Elastic cuts through our LET data in the $(0,K,L)$, $(H,K,1)$ and $(H,1,L)$ planes are shown in Fig.~\ref{fig:ElasticAlignment}. The match between the nuclear Bragg peaks and the reciprocal-lattice grid reflects the accuracy of our crystal alignment and refined structural parameters. This demonstration is important to determine the monoclinic angle $\beta$ accurately; this angle enters the neutron scattering cross-section in the dipole projection factor and hence impacts the magnetic intensity resulting from integrating the signal in the out-of-plane direction.

\section{Inelastic Neutron Scattering}
\label{ss2}

\begin{figure}[t]
    \includegraphics[width=\columnwidth]{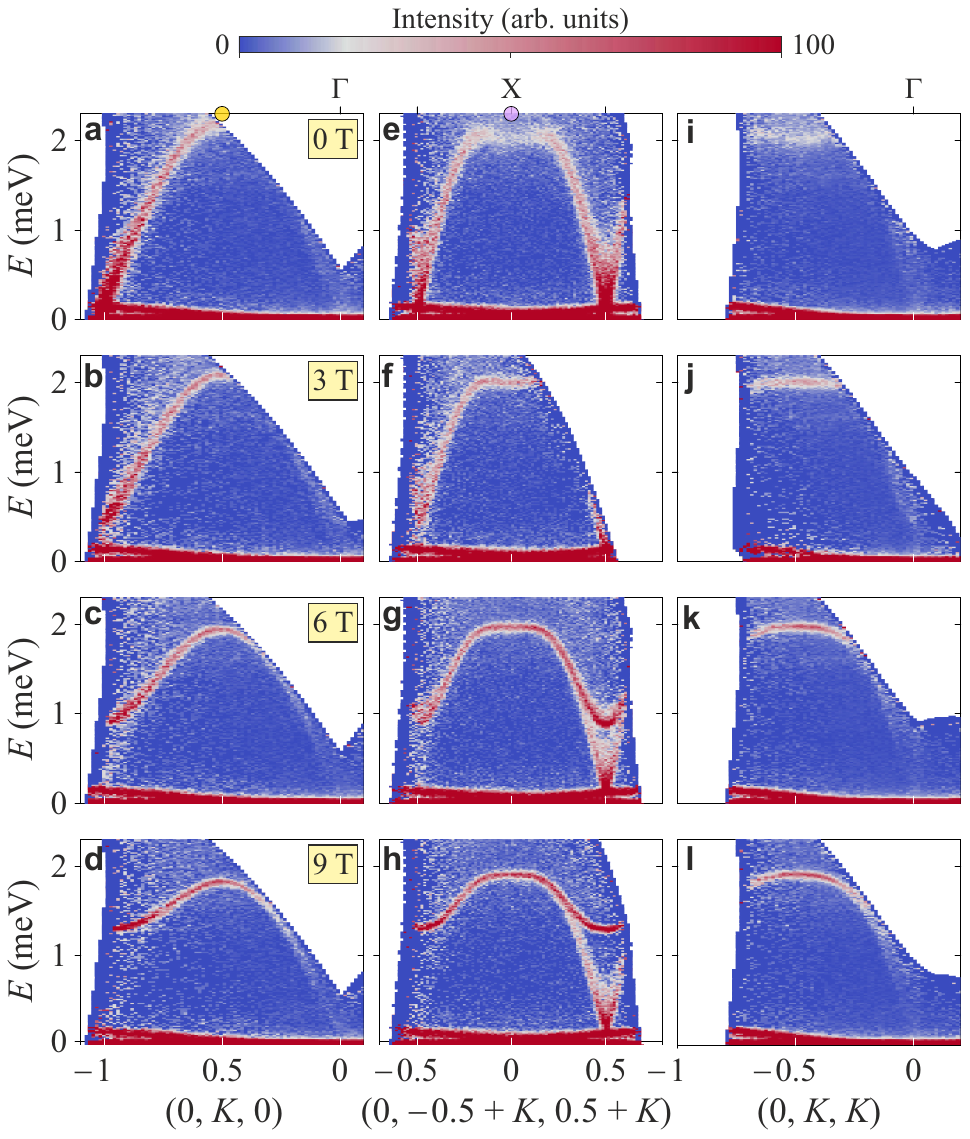}
    \caption{{\bf Field-dependence of magnetic excitations measured on LET with $E_i = 2.67$ meV.} Momentum-energy slices through the data are shown with the momentum transfer $K$ (measured in r.l.u.) varying along {\bf a-d} $(0, K, 0)$, {\bf e-h} $(0,-0.5+K,0.5+K)$, and {\bf i-l} $(0,K,K)$. } 
    \label{fig:2p67meV_LET_Cuts}
    \end{figure}

\begin{figure}[t]
    \includegraphics[width=\columnwidth]{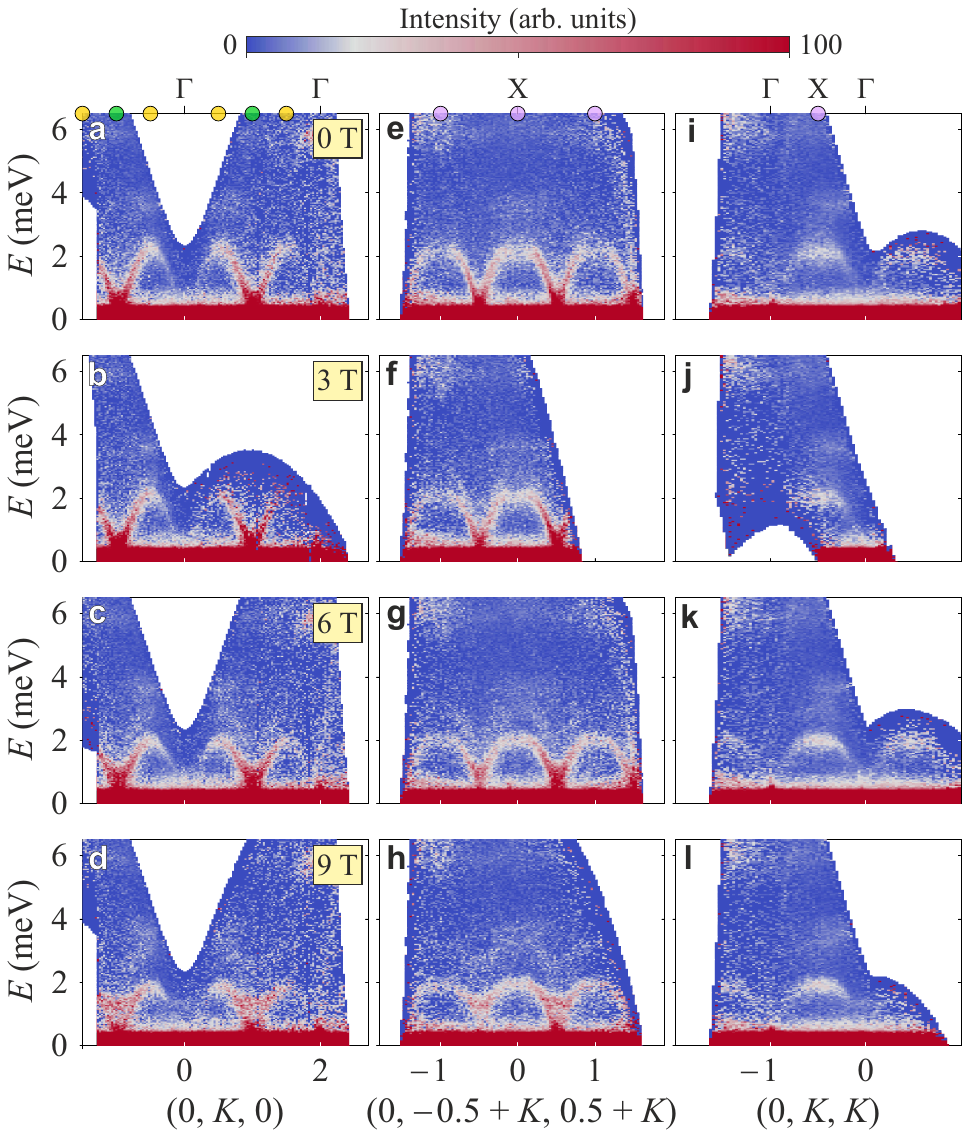}
    \caption{{\bf Field-dependence of magnetic excitations measured on LET with $E_i = 13.85$ meV.} Momentum-energy slices shown along the same directions as in Fig.~\ref{fig:2p67meV_LET_Cuts}.} 
    \label{fig:13p8meV_LET_Cuts}
\end{figure}

In our LET experiment we chose the three incident energies $E_i = 2.67$, 5.50 and 13.85 meV in order to explore the full bandwidth of the magnetic excitations in CuF$_2$(D$_2$O)$_2$(pyz). In the main text we focused on our $E_i= 5.50$~meV data, which provided the best compromise between energy resolution and energy range. For completeness, here we present further analysis of our $E_i = 2.67$ and 13.85 meV measurements, highlighting the field-dependence of selected momentum-energy slices in Figs.~\ref{fig:2p67meV_LET_Cuts} and \ref{fig:13p8meV_LET_Cuts}.

It is evident in Fig.~\ref{fig:2p67meV_LET_Cuts} that the momentum-energy coverage at  $E_i = 2.67$ meV is significantly smaller than that at $E_i = 5.50$ meV, making this incident energy too low to capture the Larmor-shadow mode (LSM).  However, it is sufficient to capture the entire one-magnon bandwidth, and indeed with a FWHM energy resolution significantly sharper than for $E_i = 5.5$ meV. Nevertheless, we find that the one-magnon excitations remain resolution-limited. 

Turning to $E_i = 13.85$ meV, the considerably larger momentum-energy coverage allows us to follow the evolution of all spectral features, including the LSM, through multiple Brillouin zones. Figure \ref{fig:13p8meV_LET_Cuts} confirms that this mode is a ubiquitous feature of the dynamical response. This dataset also illustrates the absence of significant phonon contributions below $E \approx 5$~meV, whereas the mode around $E = 6$~meV, whose intensity increases with momentum transfer for the $(0,K,K)$ and $(0,0,K)$ slices, is an optical phonon.

\begin{figure*}[t]
    \includegraphics[width=\textwidth]{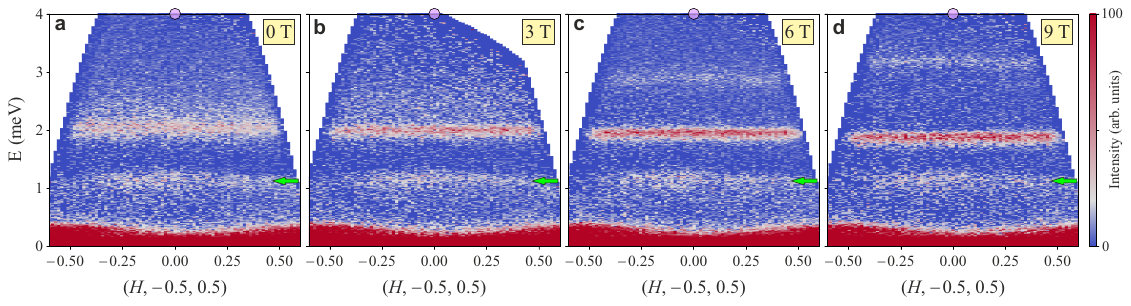}
    \caption{{\bf Out-of-plane dispersion of magnetic excitations measured on LET with $E_i = 5.50$ meV.} Field-dependence of flat excitations along the $(H,0,0)$ direction, centred around the $(0,-0.5,0.5) \equiv (\pi,0)$ point of the square-lattice Brillouin zone of CuF$_2$(D$_2$O)$_2$(pyz). Green arrows at $E = 1.2$~meV indicate again the field-independent spurious scattering signal that is present in our $E_i = 5.50$ meV data (Fig.~2 of the main text) but absent at $E_i = 2.67$ meV (Fig.~\ref{fig:2p67meV_LET_Cuts}) and $13.8$ meV (Fig.~\ref{fig:13p8meV_LET_Cuts}).}
    \label{fig:Pi0_adisp}
\end{figure*}

Measurements with variable $E_i$ also confirm our statement that the band of intensity around 1.2 meV in Fig.~2 of the main text is a consequence of scattering from the LET magnet. It is simply absent at lower $E_i$, where the low-energy signal visible in Fig.~\ref{fig:2p67meV_LET_Cuts} that seemingly disperses from the elastic line is a different type of spurion that arises from multiple scattering within the magnet. At higher $E_i$, the reflected neutrons arrive more quickly at the detector and the spurion appears at a low energy indistinguishable from the elastic line. We reiterate that the invariance of this feature with changing applied field and its absence in our TASP data also mark it unambiguously as a spurion.

\begin{figure}[t]
    \includegraphics[width=0.8\columnwidth]{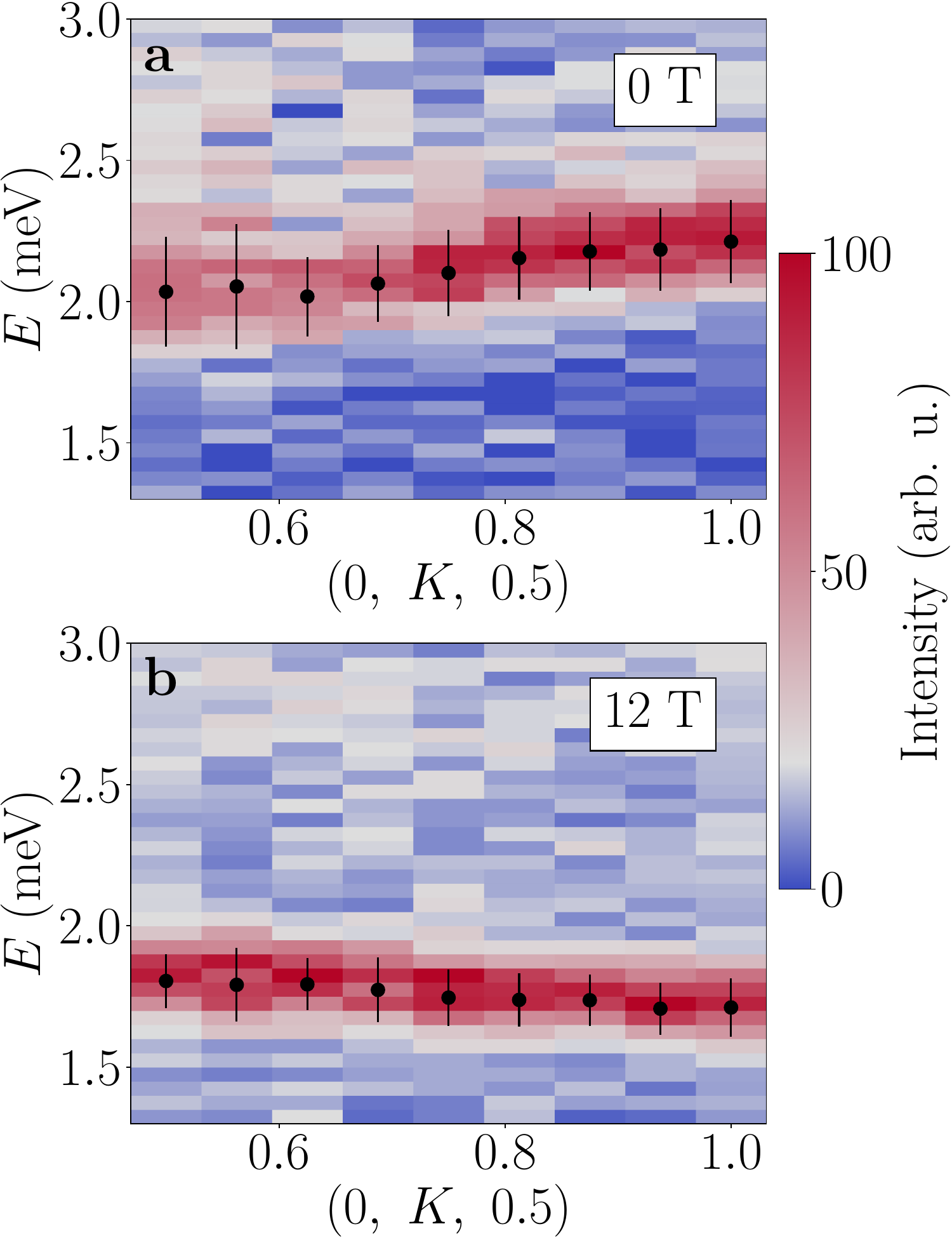}
    \caption{{\bf Field-dependence of zone-boundary excitations measured on TASP.}  {\bf a} One-magnon dispersion along the zone boundary from $(0, 0.5, 0.5) \equiv (\pi,0)$ to $(0, 1, 0.5) \equiv (\pi/2,\pi/2)$ at 0 T. {\bf b} Zone-boundary one-magnon dispersion at 12 T.} 
    \label{fig:TASP_Scans}
\end{figure}

To investigate the out-of plane dispersion (i.e.~along the $a^*$ direction) we prepared additional slices through the $E_i = 5.50$~meV dataset to focus specifically on $(H, -0.5, 0.5)$, which is centred on the zone-boundary $(\pi,0)$ point in square-lattice units. As Fig.~\ref{fig:Pi0_adisp} makes clear, both the zone-boundary one-magnon mode and the LSM have minimal dispersion at all magnetic fields measured. This is particularly clear at 6~T and 9~T, where the energies remain completely constant despite the momentum transfer spanning an entire Brillouin zone. These results justify our assumption that the magnetic interactions in CuF$_2$(D$_2$O)$_2$(pyz) are strictly two-dimensional.

\begin{figure}[t]
    \includegraphics[width=0.8\columnwidth]{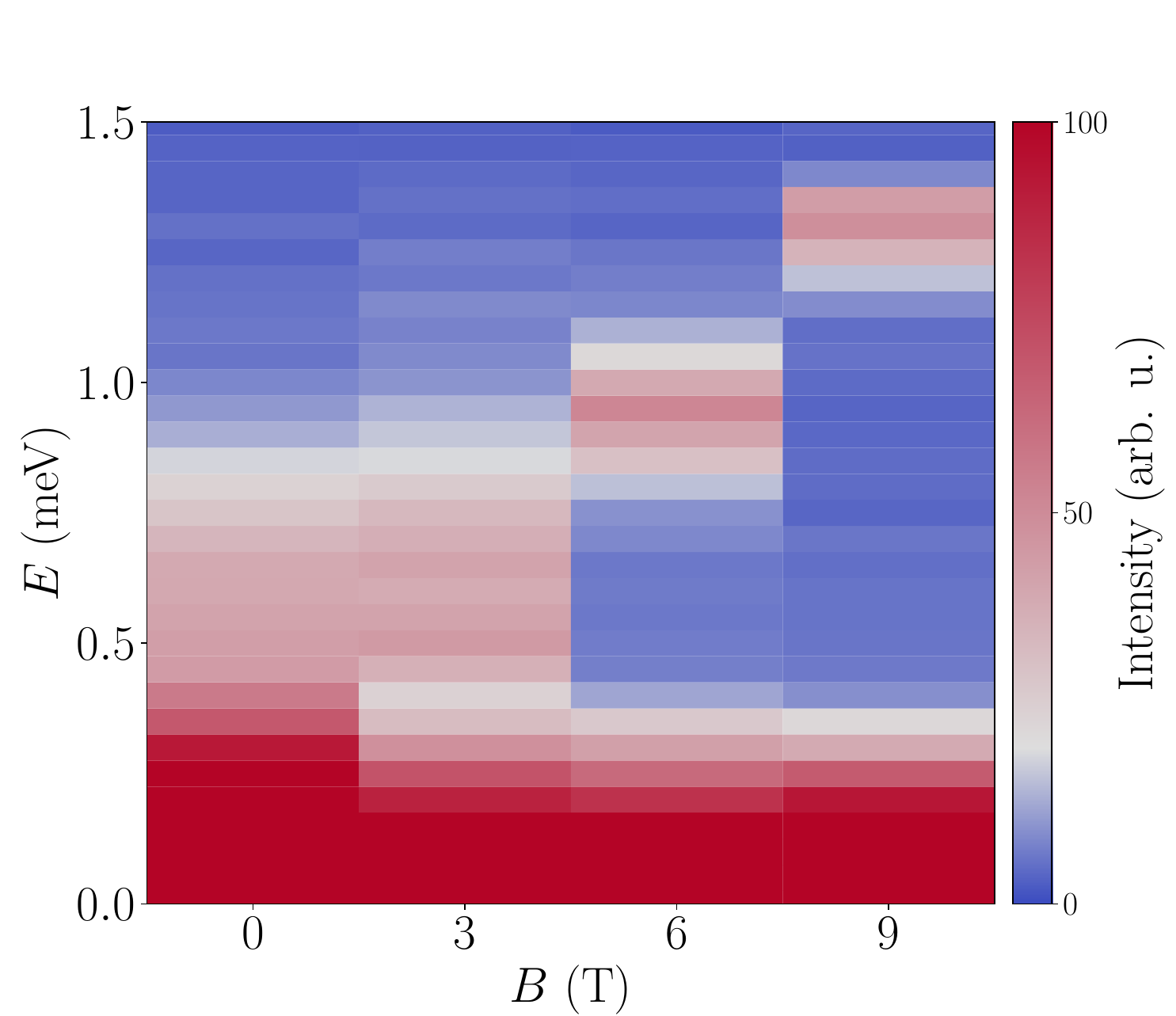}
    \caption{{\bf Field-dependence of the Larmor mode.} Energy of the gapped one-magnon branch at the M point, shown as a function of the magnetic field applied in our LET experiment. This gap measures the Larmor energy, $E_{\rm L} = g \mu_{\rm B} \mu_0 B$.} 
    \label{fig:Larmor-Mode}
\end{figure}

Figure \ref{fig:TASP_Scans} complements Fig.~4 of the main text, which showed the detailed dependence on magnetic field of the spectra measured on TASP at $(0, 0.5, 0.5) \equiv (\pi,0)$ and $(0, 1, 0.5) \equiv (\pi/2,\pi/2)$, by presenting momentum scans performed at 0 T and 12 T that connect these points. It is clear that a magnetic field of 12 T causes a significant decrease in the energy of the zone-boundary one-magnon branch, and also a significant sharpening of this branch. The reduction in linewidth is present not only at $(\pi,0)$, where increasing magnetic field eliminates the scattering anomaly, but in a less pronounced form for all momenta along the zone boundary. The LSM is not visible in these TASP datasets because it has not formed at 0~T and lies outside the available energy range at 12~T.

Figure \ref{fig:Larmor-Mode} complements Fig.~2 of the main text, by presenting the energy of the gapped mode at the M point $(0,-1,0)$ as a function of the magnetic field. Because of the unit-cell doubling in the ordered phase, this gap constitutes a measurement of the Larmor mode, and a fit to the linear field-dependence allows us to extract a $g$-factor for this field direction of $g_a \approx 2.4$.

\section{Cylinder Matrix-Product-States Calculations}
\label{ss3}

Matrix-product states (MPS) provide an Ansatz that allows a highly efficient representation of the quantum many-body states of a 1D Hamiltonian. The MPS is formulated using rank-three tensors at each lattice site and the accuracy of the Ansatz is determined by the bond dimension, $\chi$. MPS methods can be extended to 2D systems by wrapping the lattice in a cylindrical geometry, keeping periodic boundary conditions in the wrapping direction and open boundary conditions at the edges of the cylinder. Calculation of the dynamical spin structure factor (DSSF) proceeds in two steps, (i) determining the ground state of the Hamiltonian and (ii) time-evolving the system by applying the appropriate spin operators to this state. The ground state is obtained using the density-matrix renormalization group (DMRG), which is an iterative variational algorithm. Time-evolution of the state resulting from the action of a spin operator on the ground state is effected by the time-dependent variational principle (TDVP). The two-site variant of the TDVP algorithm trotterizes the projectors to the tangent space of the MPS for every adjacent pair of sites, and applying these projectors results in an effective two-site Hamiltonian. The exponential of the effective Hamiltonian ($e^{- i \frac{1}{2} H_{\mathrm{eff}} \delta t}$) acts on the MPS to give the local time-evolution, which is iterated over all sites from the left to the right end of the MPS and then back. The result is the new state time-evolved by one time step, $\delta t$. The final time to which the state is evolved is denoted by $t_f$ and the number of steps by $N$ (i.e.~$t_f = N \delta t$). 

\subsection{Ground-state order}

We perform our calculations in the laboratory frame $(x_0, y_0, z_0)$ defined in the main text, where the external magnetic field is aligned in the $z_0$ direction. This allows us to distinguish the component $\mathcal{S}^{z_0z_0}$ of the DSSF involving spin fluctuations parallel to the field, denoted $\parallel \! B$, from the components perpendicular to the field ($\perp \! B$). In order to reproduce both the INS results and the known ordered moment of the SLHAF, $m_0 = 0.61 \mu_0$, we include a small symmetry-breaking staggered field in the $x_0$ direction, and hence model the spin Hamiltonian 
\begin{eqnarray}
    H = J \! \sum_{\langle i,j\rangle} \! \mathbf{S}_{i} \! \cdot \! \mathbf{S}_{j} \! + \! h_{z} \!\! \sum_{i,j} \! S^{z_0}_{i,j} \! + \! h_{x} \!\! \sum_{i,j}, \! {(-1)}^{i+j}S^{x_0}_{i,j},
    \label{Eq:S1}
\end{eqnarray}
where $J$ is the nearest-neighbour spin-spin interaction, $h_{z}$ the external magnetic field and $h_{x}$ the staggered field. To access the complete spectra for the two different high-symmetry lines in the Brillouin zone shown in Fig.~2 of the main text, we wrapped the square lattice onto the cylinder in two different ways, namely horizontal and diagonal (i.e.~a ${45}^{\circ}$ tilt of the lattice). MPS calculations were performed at the magnetic fields corresponding to the LET measurements, $B = 0$, 3, 6 and 9~T, using a cylinder of length $L = 100$ and circumferences of $w = 4$ and 6, with bond dimension $\chi = 400$. 

The staggered field required to obtain the staggered and uniform magnetizations computed by ED and by QMC simulations\cite{Lauchli2009} within a tolerance of $10^{-2}$ was determined to be $h_x = 0.01J$. As expected, the presence of this tiny field made no difference to the excitation energies (within the effective energy resolution), but led to a discernible reduction in the spectral intensity of the one-magnon branch in the $\mathcal{S}^{x_0x_0}$ channel, with a concomitant increase in $\mathcal{S}^{y_0y_0}$; after multiplication of each with their respective polarization factors we reproduce the results measured by INS. 

\subsection{Dynamical spin structure factor}

Because the magnetic order of the ground state is uniform in $\langle S^{z_0}\rangle$ and staggered in $\langle S^{x_0}\rangle$, the magnetic unit cell contains $n_c = 2$ sites. We compute the time-dependent spin-spin correlations starting from each of these sites and take the Fourier transform
\begin{eqnarray}
    \label{eq:DSSF_defn}
    \mathcal{S}^{\alpha \beta}(\mathbf{k}, \omega) = \frac{1}{n_c} \sum_{a,b}e^{-i \mathbf{k} \cdot \left( \mathbf{r}_a - \mathbf{r}_b\right)} S^{\alpha \beta}_{a,b}(\mathbf{k},\omega),
\end{eqnarray}
with $\mathbf{r}_a$ and $\mathbf{r}_b$ the relative positions of sites $a$ and $b$ within the unit cell and 
\begin{eqnarray}
    S^{\alpha \beta}_{a,b} (\mathbf{k},\omega) & = & \frac{1}{N_c} \sum_{\mathbf{R}} \int_{-\infty}^{\infty} e^{i\left( \omega t - \mathbf{k} \cdot \mathbf{R} \right)} C^{\alpha,\beta}_{\mathbf{R}, \mathbf{0}}(t) dt, \\
    C^{\alpha,\beta}_{\mathbf{R}, \mathbf{0}}(t) & = & \langle S^\alpha_{a, \mathbf{R}}(t) S^{\beta}_{b,\mathbf{0}}(0) \rangle - \langle S^\alpha_{a, \mathbf{R}}(0) \rangle \langle S^{\beta}_{b,\mathbf{0}}(0) \rangle, \nonumber
\end{eqnarray}
where $N_c$ is the number of unit cells, $\mathbf{R}$ the spatial position of each cell and $\alpha = \beta \in \lbrace x_0, y_0, z_0\rbrace$ are the spin components. 
The spread of the time-dependent spin-spin correlations limits the maximum time up to which a state can be evolved. The long cylinders we choose set an upper limit of $t_f = 32/J$ before a spin operation at the centre of the cylinder can reach the open boundaries. Setting this cut-off prevents Friedel oscillations, which would introduce uncontrolled numerical artifacts in the spectra. To avoid the Gibbs oscillations that arise due to the finite evolution time, it is standard to multiply $C^{\alpha,\beta}_{\mathbf{R}, \mathbf{0}}(t)$ by a Gaussian filter in time when performing the Fourier transform. This filter function is expressed as  
$$
f(t) = \frac{1}{\sqrt{2\pi \sigma_t^2}} e^{- t^2/2\sigma_t^2},
$$
where we set $\sigma_t = \tfrac{1}{2} t_f$. 

\begin{figure}[t]
\includegraphics[width=0.99\columnwidth]{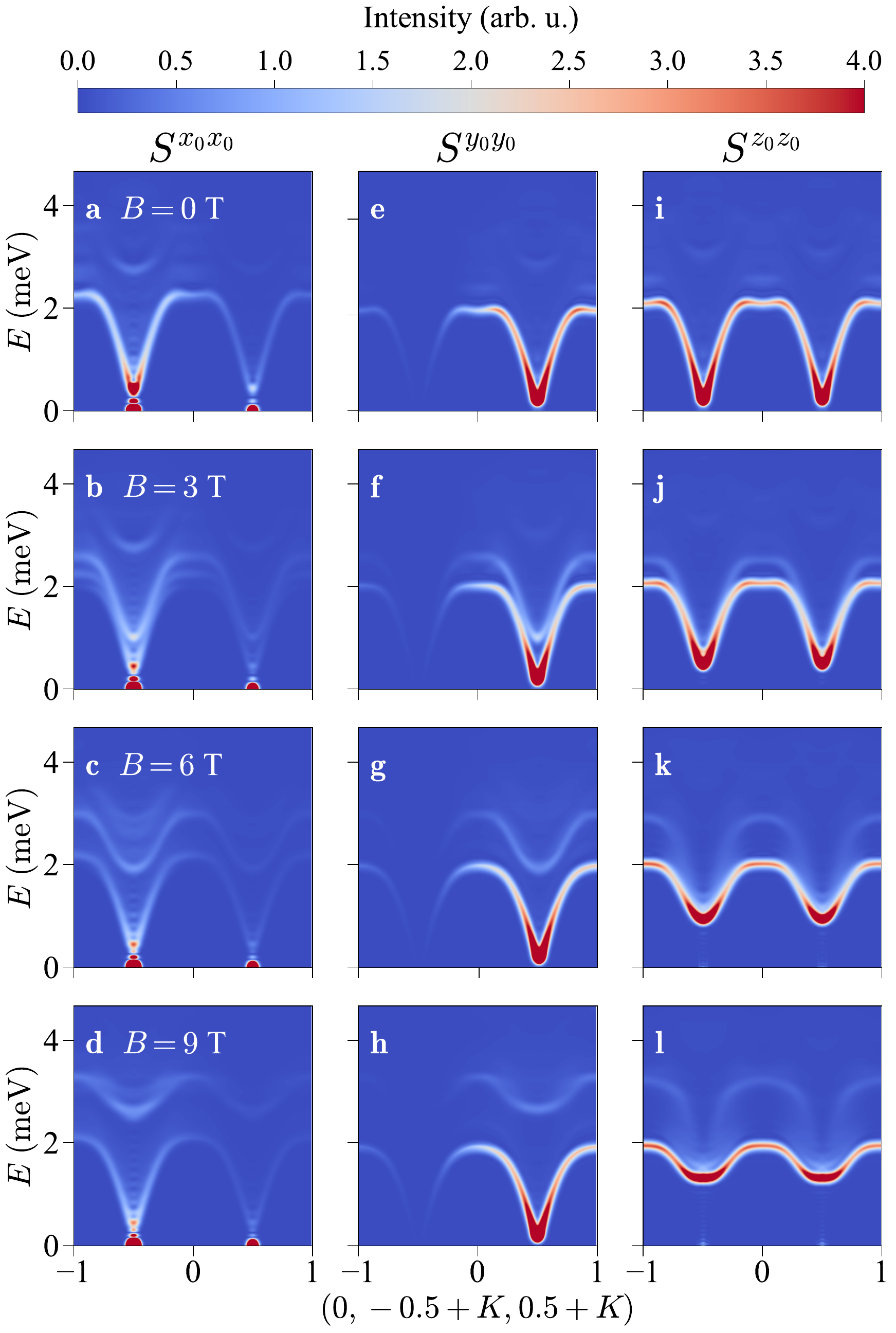}
\caption{{\bf DSSF along MXM calculated by MPS for the three separate spin channels.} The DSSF components $\mathcal{S}^{x_0x_0}$ (a-d), $\mathcal{S}^{y_0y_0}$ (e-h) and $\mathcal{S}^{z_0z_0}$ (i-l) with polarization factor are shown for the four magnetic fields of the LET experiment and for the $(0, -0.5+K, 0.5+K)$ direction.
\label{fig:MPS_Pol_Direction1}}
\end{figure}

\subsection{Spin channels and polarization factor}

The external magnetic field fixes the direction $z_0$ (corresponding to the crystalline $a$ axis) and the staggered field is applied in the $x_0$ direction, meaning that the ordered moments have no component along $y_0$, which then corresponds to the ${c}^\ast$ axis. The polarization factor in the INS cross-section modulates the components of the DSSF according to
\begin{eqnarray}
P_\alpha(\mathbf{Q}) = \left(1 - \frac{Q_\alpha^2}{\mathbf{Q}^2}\right), 
\end{eqnarray}
where $\mathbf{Q} = H{\it {\bf a}}^* +K{\it {\bf b}}^* + L{\it {\bf c}}^*$ and the components are expressed in the laboratory frame, meaning that $\alpha \in \lbrace x_0, y_0, z_0\rbrace$, corresponding to the $(c^\ast, - b^\ast, a)$ directions in the crystal. The total INS cross section then satisfies
\begin{eqnarray}
\label{eq:total_cross_section}
I(\mathbf{Q},\omega) \propto \sum_{\alpha} P_{\alpha} (\mathbf{Q}) S^{\alpha\alpha} (\mathbf{Q},\omega),
\end{eqnarray}
where $S^{\alpha\alpha} (\mathbf{Q},\omega)$ is defined in Eq.~\eqref{eq:DSSF_defn}.

\begin{figure}[t]
\includegraphics[width=0.99\columnwidth]{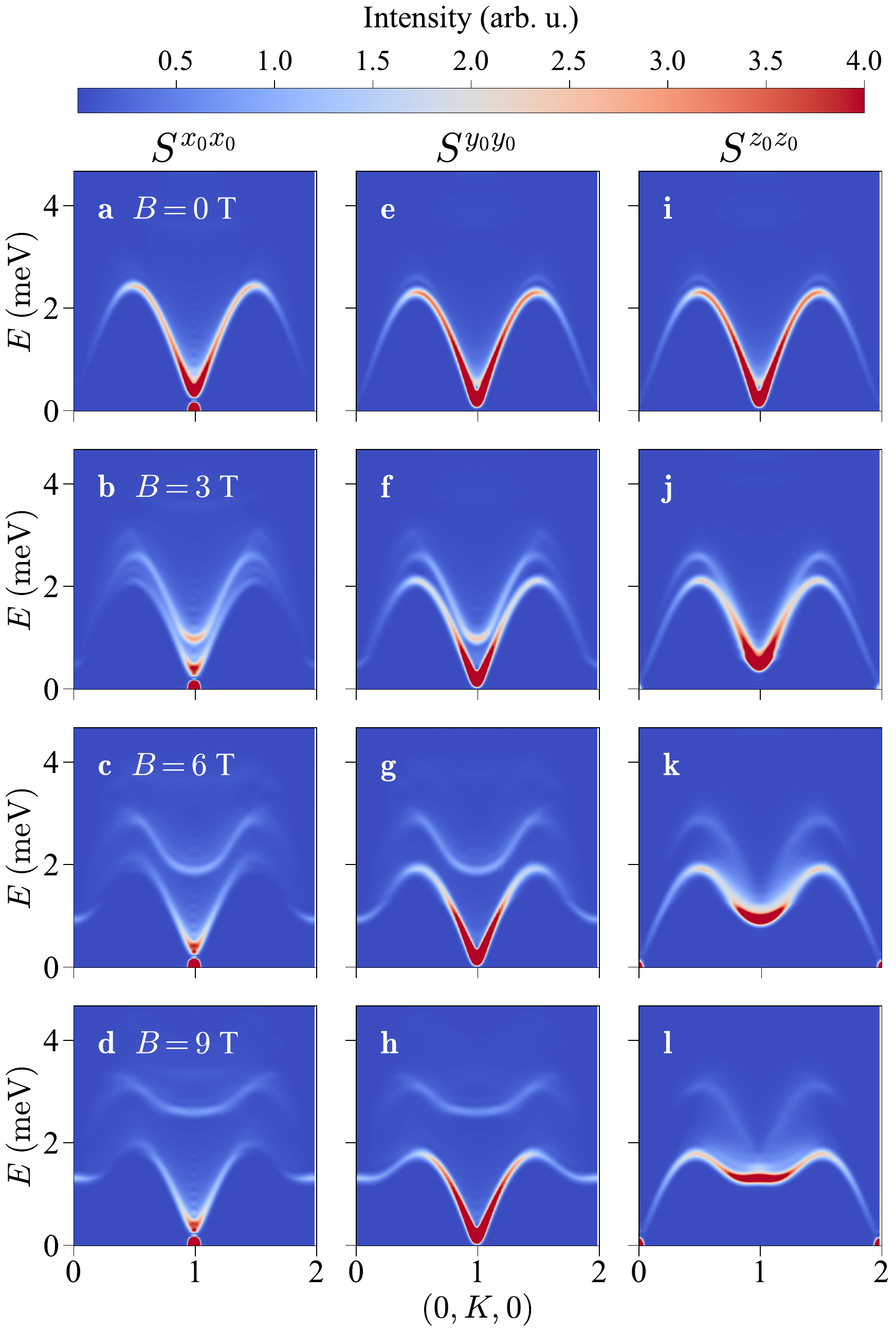}
\caption{{\bf DSSF along $\Gamma$M$\Gamma$ calculated by MPS for the three separate spin channels.} The DSSF components $\mathcal{S}^{x_0x_0}$ (a-d), $\mathcal{S}^{y_0y_0}$ (e-h) and $\mathcal{S}^{z_0z_0}$ (i-l) are shown for the four magnetic fields of the LET experiment and for the $(0,K,0)$ direction (Figs.~2(a-h) of the main text).
The polarization factor is not included because it is uniformly unity for $\mathcal{S}^{x_0x_0}$ and $\mathcal{S}^{z_0z_0}$, and uniformly zero for $\mathcal{S}^{y_0y_0}$.
\label{fig:MPS_Pol_Direction2}}
\end{figure}

\begin{figure*}[t]
\includegraphics[width=0.84\textwidth]{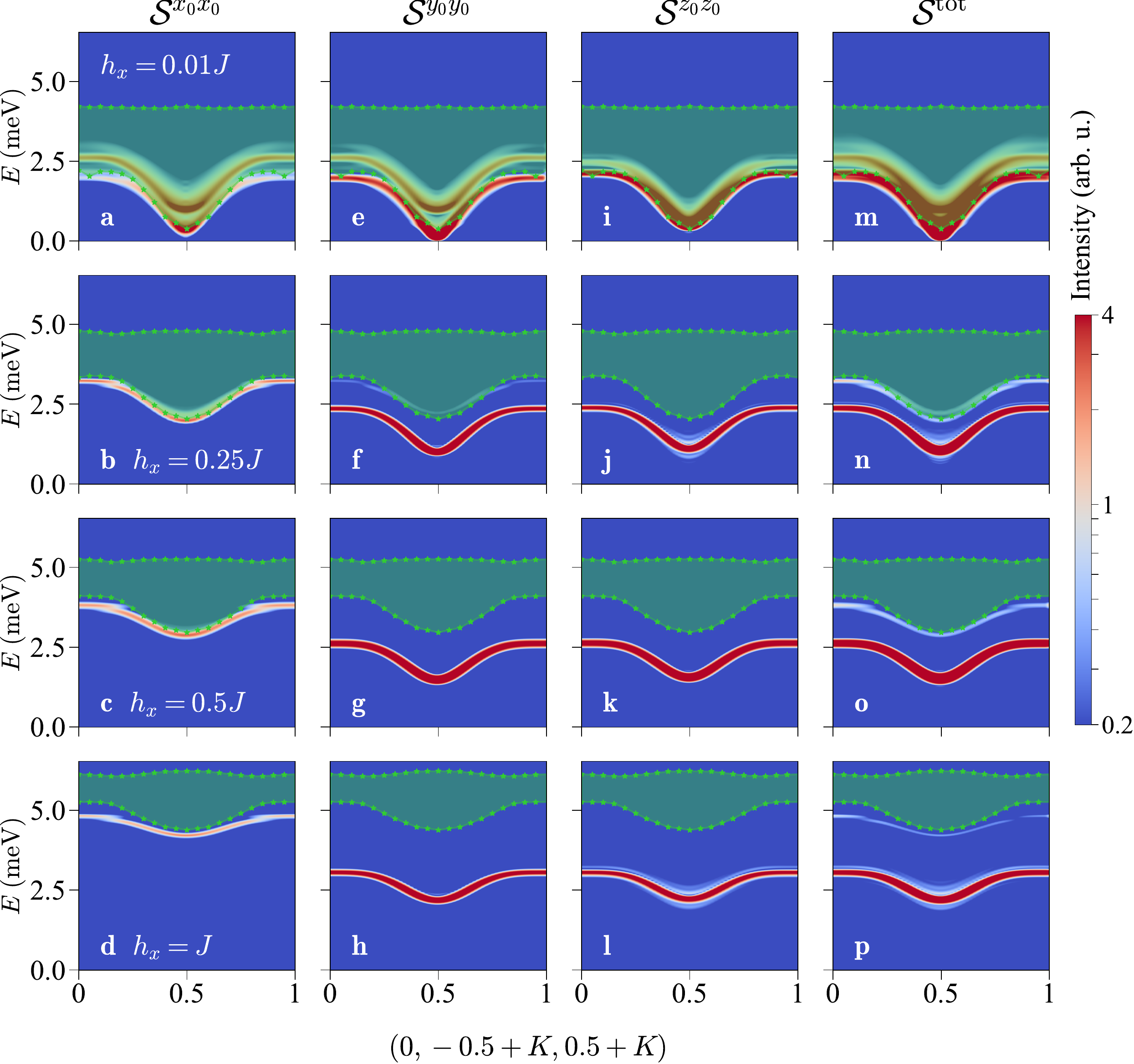}
\caption{{\bf Characteristics of the LSM at different staggered magnetic fields.} DSSF components $S^{x_0x_0}$ (a-d), $S^{y_0y_0}$ (e-h), $S^{z_0z_0}$ (i-l) and their sum combined with the INS polarization factor, denoted $S_{\rm tot}$ (m-p), shown for the XMX direction (specifically, $(0,-0.5+K,0.5+K)$) at different $h_x$ values and a fixed longitudinal field of $B = 3$~T. We use a logarithmic colour scale to highlight the LSM. The magnitude of staggered field is increased from top to bottom, taking the values $h_x = 0.01J$ (a,e,i,m), $0.25J$ (b,f,j,n), $0.5J$ (c,g,k,o) and $J$ (d,h,l,p). The two-magnon continuum is represented by the green shading (based on the boundary points represented by the small stars). The bound state is visible most prominently in the $S^{x_0x_0}$ channel and lies outside the two-magnon continuum for the full range of $k_x$ at $h_x = J$. } 
\label{fig:staggeredfield}
\end{figure*}

The separate spin channels $\mathcal{S}^{x_0x_0}$, $\mathcal{S}^{y_0y_0}$ and $\mathcal{S}^{z_0z_0}$ obtained from MPS are shown in Fig.~\ref{fig:MPS_Pol_Direction1} for the $(0, -0.5+K, 0.5+K)$ direction, which corresponds to the experimental measurements shown in Figs.~2i-l of the main text, while the sum of these three components is shown in Figs.~2m-p. We observe that each channel contains only one magnon branch and one LSM, but that these are not of the same character, with the exact $E_{\rm L}$-shifted copy of a given branch in $\mathcal{S}^{x_0x_0}$ and $\mathcal{S}^{y_0y_0}$ (which are the same up to polarization factors) appearing in $\mathcal{S}^{z_0z_0}$ and conversely. In Fig. \ref{fig:MPS_Pol_Direction2} we separate the spin channels for the DSSF in the $(0, K, 0)$ direction, which corresponds to the measurements shown in Figs.~2a-d of the main text, with the sum of the three components in Figs.~2e-h. Here we do not include the polarization factor, which is independent of $K$; this factor multiplies the $\mathcal{S}^{y_0y_0}$ channel by zero, such that it makes no contribution to the DSSF, but we show it for for completeness. Again we observe the same branches in $\mathcal{S}^{x_0x_0}$ and $\mathcal{S}^{y_0y_0}$ and the channel-switching nature of the processes forming the two LSMs. 

We stress that, when comparing our MPS results with INS in Fig.~2 of the main text, the same scale factor is used at all four applied magnetic fields. We remark also that, in making this type of comparison, we have not modelled the instrumental resolution function in detail, but instead choose the width of the Gaussian filter in our MPS calculations (Sec.~\ref{ss3}B) to achieve an approximate linewidth match; again only one value of $\sigma_t$ is used throughout. These comparisons provide a complete validation of the SLHAF as the model and for the orientation of the canted magnetic structure with respect to the crystal axes, specifically that they rotate in the $ac^\ast$ plane as the field is increased. 

\subsection{Nature of the Larmor-Shadow Mode}

In the main text we showed that the primary hallmarks of the LSM observed by INS and MPS are its shift by the Larmor energy from the one-magnon branch and its very sharp (albeit intrinsic) linewidth. However, this mode lies within the boundaries of the conventional two-magnon continuum, and hence it is expected to be an anomalously narrow resonance, or quasi-bound state, rather than being a true bound state. One way to investigate how the resonance we observe is related to a true bound state is to increase the staggered-field parameter ($h_x$) in our MPS calculations based on the Hamiltonian of Eq.~\eqref{Eq:S1}. 

The first step of this analysis is to determine the boundaries of the two-magnon continuum. A robust calculation of the one-magnon dispersion on a dense grid of evenly spaced momenta in ($k_x,k_y$) would lead directly to the continuum boundaries for two noninteracting magnons. An unbiased calculation of the magnon dispersion can be effected by following the intensity maximum in the MPS DSSF. However, the finite circumference of the cylinder restricts this exercise to a limited number of $k_y$ values, specifically $k_y = 0$, $\pi/2$, $\pi$ and $3\pi/2$ for $w = 4$ and in addition $k_y = \pi/3$, $2\pi/3$, $4\pi/3$ and $5\pi/3$ with $w = 6$. To obtain an adequately dense grid in $k_y$, we interpolate between the available values over the range from 0 and $2\pi$ by describing the magnon dispersion as a sum of cosines, 
$$
\omega(k_x,k_y) = \sum_{n=0}^{5} a_n (k_x) \cos(n k_y),
$$
fitted to the DSSF data. The upper and lower boundaries of the two-magnon continuum are then given by
\begin{eqnarray}
\omega_{\mathrm{u}} (K_x, K_y) & = & \max [\omega(k_{x_1},k_{y_1}) + \omega(k_{x_2} + \pi, k_{y_2} + \pi)], \nonumber \\
\omega_{\mathrm{l}} (K_x, K_y) & = & \min [\omega(k_{x_1},k_{y_1}) + \omega(k_{x_2} + \pi, k_{y_2} + \pi)], \nonumber
\end{eqnarray}
where $K_x = k_{x_1} + k_{x_2}$ mod 2$\pi$ and $K_y = k_{y_1} + k_{y_2}$ mod 2$\pi$. We estimate the uncertainties in this procedure from the fact that width-4 and width-6 cylinders share two wavevectors, and the differences in magnon energies at these two points lie within $0.025J$ for all values of $k_x$. These differences are finite-size effects, and taking the uncertainty in the magnon dispersion relation to be $0.025J$ sets the error bars on the upper and lower boundaries of the two-magnon continuum as $0.05J$. 

Figure \ref{fig:staggeredfield} shows the separate spectral functions in all three spin channels for four representative values of $h_x$. Considering first the one-magnon branches, the $h_x$ value we used to model CuF$_2$(D$_2$O)$_2$(pyz) (Figs.~\ref{fig:staggeredfield}a,e,i) has no effect other than to remove the U(1) rotational symmetry. Increasing $h_x$ causes the progressive opening of a full gap, a flattening of the bandwidth and an increase in the band centre, which together cause a progressive raising and narrowing of the two-magnon continuum. By comparing its location to the boundaries of this continuum, one may trace the evolution of the LSM in the $\perp \! B$ spin channels from a rather sharp resonance within the continuum at small $h_x$ to a well defined (resolution-limited) and separate mode lying completely below the continuum as $h_x$ approaches $J$. 

In that limit, $h_x$ dominates both $B$ and $J$ to enforce a ground state with staggered spin orientation in the $x_0$ direction. One-magnon excitations can arise only due to the action of $S^{y_0}$ or $S^{z_0}$ in flipping the spins of this state, and hence appear only in the $S^{y_0y_0}$ and $S^{z_0z_0}$ channels. The bound state is a distinct and stable mode arising from two spin-flips that are adjacent, giving them an energy $J$ lower than two independent spin-flips. Although this mode appears in both $S^{x_0x_0}$ and $S^{y_0y_0}$, the absence of an intense one-magnon branch in $S^{x_0x_0}$ makes it straightforward to track the evolution of the bound state into a resonance with decreasing $h_x$ in this channel.  

\section{Spin-Wave Theory}
\label{ss4}

Spin-wave theory (SWT) provides a systematic framework in which to interpret the properties of ordered magnetic systems. Despite the technical complexities inherent to any expansion beyond the lowest orders, this framework nevertheless affords considerable insight into the microscopic spin-fluctuation processes underlying many of the physical properties displayed even by systems with strong quantum fluctuations. With a view to obtaining deeper insight into the origin of the LSM, we applied SWT including $1/S$ corrections to calculate the zero-temperature DSSF of the quasi-2D square-lattice Heisenberg antiferromagnet (SLHAF) following the approach of Ref.~\onlinecite{Fuhrman2012} It is straightforward to adapt this calculation to the geometry of our CuF$_2$(D$_2$O)$_2$(pyz) experiments and hence  to compare its output with our INS and cylinder MPS results. 

\subsection{General geometrical considerations}

We consider the Hamiltonian
\begin{eqnarray}
    {H} = J \sum_{\langle ij \rangle} \mathbf{S}_i \! \cdot \! \mathbf{S}_j + J' \! \sum_{\langle ij \rangle_z} \mathbf{S}_i \! \cdot \! \mathbf{S}_j - g \mu_B B \sum_{i} \! S_i^{z_0}, 
    \label{eq:ham}
\end{eqnarray}
with the magnetic field  applied along the $z_0$ direction in the laboratory frame ($z_0 \equiv [100]$ matches the $a$ axis in our sample alignment, with the $b^\ast c^\ast$ plane horizontal). Here $g$ is the gyromagnetic ratio of the Cu$^{2+}$ magnetic moments in the field direction and in addition to the nearest-neighbour $J$ in the crystallographic $bc$ plane we consider an interplane interaction $J'$ parametrized by $\alpha = J'/J$ ($0 \leq \alpha \leq 1$). Both interactions are antiferromagnetic and $\alpha$ in CuF$_2$(D$_2$O)$_2$(pyz) is demonstrably very small, but sufficient to stabilize long-range magnetic order at the N\'eel temperature, $T_{\rm N} = 2.6$~K.

\begin{figure}[t]
\includegraphics[width=0.6\columnwidth]{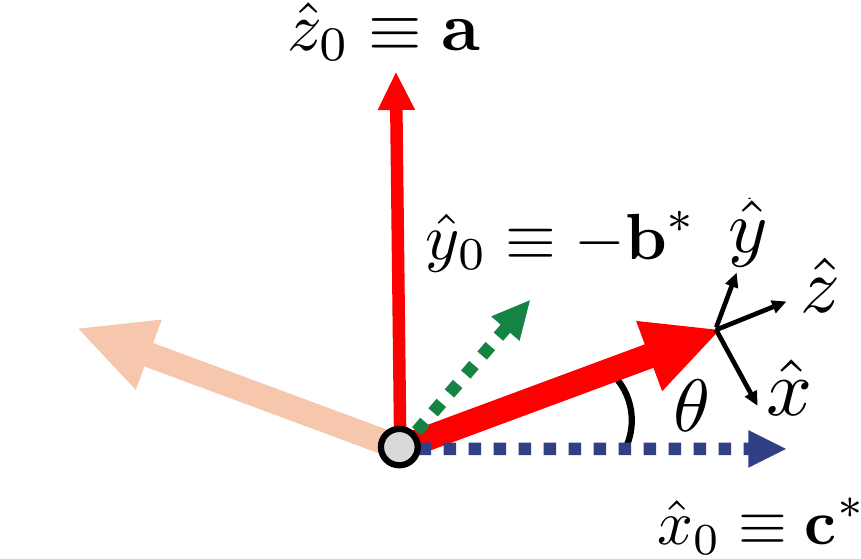}
\caption{{\bf Canted magnetic structure.}  Depiction of the assumed magnetic order within the laboratory frame, $(\hat{x}_0, \hat{y}_0, \hat{z}_0)$, which defines the  local frame, $(\hat{x},\hat{y},\hat{z})$.} 
\label{fig:Frames}
\end{figure}

To calculate the spin components in the Cartesian laboratory frame $(\hat{x}_0,\hat{y}_0,\hat{z}_0)$, we assume a canted antiferromagnetic order for any $B > 0$, as shown in Fig.~\ref{fig:Frames}. This order is associated with the propagation vector $\mathbf{Q}_m = (\pi,\pi,\pi)$ and with a field-dependent canting angle, $\theta$, by which the spins tilt uniformly out of the spin-flop plane, which is that perpendicular to $B$. The wavevector $\mathbf{Q}_0 = (0,0,0)$ corresponding to the uniform magnetization can be viewed as a secondary propagation vector. The U(1) symmetry of this system corresponds to one remaining degree of freedom, the spin projection in the $x_0y_0$ ($b^\ast c^\ast$) plane, which we express as $\phi$ measured from the $x_0$ axis and assume to be independent of $\theta$. 

In CuF$_2$(D$_2$O)$_2$(pyz) at zero field the spins are aligned along $[0.7,0,1]$ in real space, meaning that there is no ordered component along $\hat{b} \equiv [0,1,0]$, which coincides with $\hat{b}^\ast$. Thus we assume that, just after the spin-flop ($B = 0^+$), the spins align perpendicular to the field direction ($\hat{a}$) and perpendicular to $\hat{b}^\ast$, i.e.~that the spins point initially along ${\hat c}^\ast \equiv (0,0,1)$ and tilt gradually towards $\hat{a}$ with increasing field. To maintain a right-handed coordinate system, we set $\hat{x}_0$ along $c^\ast \equiv (0,1,0)$ ($\phi = 0$) and thus $\hat{y}_0$ along $- b^\ast \equiv (0,0,\bar{1})$. Finally, based on the parameters we use to model CuF$_2$(D$_2$O)$_2$(pyz) ($J = 0.934$~meV, $\alpha = 0.01$, $B_{{\rm sat},a} = 28.6$~T, $g_a = 2.4$), we expect $\theta$ to range from 0 to $\theta_{\rm max} \approx \arcsin(0.4) = 23.6^\circ$ at fixed $\phi = 0$.

The relationship between spin components in the local and laboratory frames is\cite{Fuhrman2012}  
\begin{eqnarray}
    {S}_i^{z_0} & = &  \sin \theta S_i^{z} - e^{i {\bf Q}_m \cdot {\bf r}_i} \cos \theta S_i^{x},  \\ 
    {S}_i^{y_0} & = & e^{i {\bf Q}_m \cdot {\bf r}_i} \cos \theta \sin \phi S_i^{z} + \cos \phi S_i^{y} + \sin\theta \sin\phi S_i^{x}, \nonumber \\
    {S}_i^{x_0} & = & e^{i {\bf Q}_m \cdot {\bf r}_i} \cos \theta \cos \phi S_i^{z} + \sin \phi S_i^{y} + \sin\theta \cos\phi S_i^{x}, \nonumber  
    \label{eq:Frame-Spins}
\end{eqnarray}
which for $\phi = 0$ yields
\begin{eqnarray}
    {S}_i^{z_0} & = & \sin \theta S_i^{z} - e^{i {\bf Q}_m \cdot {\bf r}_i} \cos \theta S_i^{x},  \\ 
    {S}_i^{y_0} & = & S_i^{y}, \nonumber \\
    {S}_i^{x_0} & = & e^{i {\bf Q}_m \cdot {\bf r}_i} \cos \theta S_i^{z} + \sin\theta S_i^{x}. \nonumber  
    \label{eq:Frame-Spins2}
\end{eqnarray}
The Fourier transform $S^\alpha_{\bf k} = (1/\sqrt{N}) \sum_i e^{-i {\bf k} \cdot {\bf r}_i} S_i^\alpha$ then leads to
\begin{eqnarray}
    {S}_{\bf k}^{z_0} & = & \sin \theta S_{\bf k}^{z} - \cos\theta S_{{\bf k} - {\bf Q}_m}^{x}, \nonumber  \\ 
    {S}_{\bf k}^{y_0} & = & S_{\bf k}^{y}, \\
    {S}_{\bf k}^{x_0} & = & \cos \theta S_{{\bf k} - {\bf Q}_m}^{z} + \sin\theta S_{\bf k}^{x}. \nonumber 
    \label{eq:Frame-Spins-FT}    
\end{eqnarray}
We define the component $\alpha\beta$ of the DSSF as
\begin{equation}
    \mathcal{S}^{\alpha\beta}({\bf k},\omega) = \int \frac{{\rm d} t}{2\pi} e^{ i \omega t} 
    \langle {S}^\alpha_{\bf k}(t) {S}^\beta_{-{\bf k}}(0) \rangle
\end{equation}
and obtain the diagonal components in the laboratory frame in the form
\begin{eqnarray}
\label{eq:Diagonal-DSSF}
     \mathcal{S}^{z_0 z_0}({\bf k},\omega) & = & \sin^2\theta \, \mathcal{S}^{{z}{z}}({\bf k},\omega) + \cos^2\theta \, \mathcal{S}^{{x}{x}}({{\bf k} - {\bf Q}_m},\omega), \nonumber \\
     \mathcal{S}^{y_0 y_0}({\bf k},\omega) & = & \mathcal{S}^{{y}{y}}({\bf k},\omega),  \\
     \mathcal{S}^{x_0 x_0}({\bf k},\omega) & = & \cos^2\theta \,  \mathcal{S}^{{z}{z}}({{\bf k} - {\bf Q}_m}, \omega) +  \sin^2\theta \, \mathcal{S}^{{x}{x}}({\bf k},\omega), \nonumber 
\end{eqnarray}
where following previous studies\cite{Zhitomirsky1999,Mourigal2010,Fuhrman2012} we have neglected components off-diagonal in the local frame.

For an explicit definition of terminology, we remark that spin fluctuations appearing in the ${x}{x}$ and ${y}{y}$ response functions are transverse to the ordered moments and typically are dominated by one-magnon excitations. In the laboratory frame, these terms appear at wavevector ${\bf k}$ in the $\perp \! B$ (in-plane) channel, while spin fluctuations in ${x}{x}$ also appear at wavevector ${\bf k}-{\bf Q}_m$ in the $\parallel \! B$ (out-of-plane) channel. The situation is reversed for spin fluctuations appearing in $zz$, which are longitudinal with the ordered moment, are usually dominated by two-magnon excitations and appear at wavevector ${\bf k}$ (${\bf k}-{\bf Q}_m$) in the $\parallel \! B$ ($\perp \! B$) channel. In the main text we defined the notation for these channels as $\mathcal{S}_{\perp B}^{{\alpha}{\alpha}}$ and $\mathcal{S}_{\parallel B}^{{\alpha}{\alpha}}$, where the superscript indicates components in the local frame and the subscripts refer to their projection in the laboratory frame.

After applying the magnetic form and projection factors, the scattered intensity at transferred momentum {\bf Q} and energy $E = \hbar\omega$ is
\begin{eqnarray}
    \mathcal{I}({\bf Q},\omega) & = & r_0^2 |f(Q)|^2 \big\{ g_a^2 \mathcal{S}^{z_0 z_0}({\bf Q},\omega)  \label{eq:neutron} \\ 
      & & + [1 - (Q_{c^\ast}/|{\bf Q}|)^2] \, g_{c^\ast}^2\mathcal{S}^{x_0 x_0}({\bf Q},\omega) \nonumber \\
      & & + [1 - (Q_{b^\ast}/|{\bf Q}|)^2] \, g_{b^\ast}^2 \,\mathcal{S}^{y_0 y_0}({\bf Q},\omega) \big\}, \nonumber
\end{eqnarray} 
where $\mathbf{Q} = H{\it {\bf a}}^* +K{\it {\bf b}}^* + L{\it {\bf c}}^*$, $|f(Q)|^2$ is the squared form factor of Cu$^{2+}$, $(g_a, g_{b^\ast}, g_{c^\ast})$ are the relevant projections of the gyromagnetic tensor and $r_0^2$ sets the units of measurement. Finally, to express the components of the DSSF in conventional reciprocal-space units, we account for the centred nature of the $b^\ast c^\ast$ plane in CuF$_2$(D$_2$O)$_2$(pyz) by defining the unit vectors of the reciprocal square lattice in terms of the crystalline reciprocal lattice as ${\bf e}_x = \tfrac{1}{2} ({\it {\bf b}}^\ast + {\it {\bf c}}^\ast)$ and ${\bf e}_y = \tfrac{1}{2} ({\it {\bf b}}^\ast - {\it {\bf c}}^\ast)$, leading to $k_x = \pi (K + L)$, $k_y = \pi (K - L)$ and $k_z = \pi H$, all of which repeat mod 2$\pi$.

\subsection{Magnon-magnon interactions}

We begin our calculations of the magnon dispersion and DSSF with a general introduction to the treatment of magnon-magnon interactions within SWT. Quite generally, interactions between magnons act to alter the quasiparticle dispersion and linewidth, causing energy renormalization, spontaneous decay and thermal decay, which lead to shifted and broadened lineshapes in the DSSF, as well as to changes in thermomagnetic quantities. These effects modify the structure and intensity of the multi-magnon continuum, including the van Hove singularities, and can in principle create quasiparticle bound states below or even inside the multimagnon continuum. 

In the SLHAF, magnon interactions originate from the expansion of Eq.~\eqref{eq:ham} in terms of Holstein-Primakoff bosons.\cite{Zhito2013} At zero field,\cite{Canali1992,Igarashi1992} the first non-trivial corrections originate from quartic and sextic terms in the expansion, defined in Ref.~\onlinecite{Fuhrman2012} as $\hat{\mathcal{H}}_4$ and $\hat{\mathcal{H}}_6$. These terms yield corrections of order $1/S^2$ (and higher) compared to linear SWT (harmonic order, ${\cal{O}}(1)$), as we summarize in Table \ref{tab:interactions}. The impact of these interactions on the zero-field one-magnon dispersion has been studied perturbatively up to order $1/S^3$,\cite{Syromyatnikov2010} which revealed an upward renormalization of magnon energies by approximately 18\% and a weak (3\%) modulation of the dispersion along the Brillouin-zone boundary (associated with the slow convergence of the $1/S$ expansion around ${\bf k} = (\pi,0)$). The zero-field two-magnon continuum, which carries sizable spectral weight due to the 40\% zero-point moment reduction, has been computed perturbatively to order $1/S^2$.\cite{Elliott1969} Beyond this, high-order numerical calculations for the multimagnon continuum,\cite{Powalski,Powalski2018} which are made challenging by the gapless nature of the system, have been shown to account for the anomalous transverse and longitudinal multimagnon lineshapes observed in INS experiments at ${\bf k} = (\pi,0)$,\cite{DallaPiazza} suggesting significant attractive magnon-magnon interactions. 

When an applied magnetic field causes spin canting,\cite{Zhitomirsky1998, Zhitomirsky1999,Mourigal2010,Fuhrman2012} the magnon-magnon interactions are dominated by cubic and quartic terms ($\hat{\mathcal{H}}_3$ and $\hat{\mathcal{H}}_4$). Nontrivial corrections to the magnon dispersion in finite field then emerge already at order $1/S$ (Table \ref{tab:interactions}). To serve as a benchmark of our INS and MPS results, here we provide a perturbative calculation of the DSSF in this (low-field) regime, and hence include only interaction terms arising from $\hat{\mathcal{H}}_3$ and $\hat{\mathcal{H}}_4$.

\begin{table}[b]
\begin{tabular}{|l|c|c|} \hline\hline
& 1st Order (HF)         & 2nd Order (SE) \\\hline\hline
Order $1/S^2$ ($B = 0$) & $\langle \mathcal{\hat{H}}_4 \rangle$  and $\langle \mathcal{\hat{H}}_6 \rangle$  & $\hat{\mathcal{H}_4}$  \\
Order $1/S^3$ ($B = 0$) & $\langle \mathcal{\hat{H}}_4 \rangle$, $\langle \mathcal{\hat{H}}_6 \rangle$, $\langle \mathcal{\hat{H}}_4 \mathcal{\hat{H}}_6 \rangle$ & $\hat{\mathcal{H}_4}$ and $\hat{\mathcal{H}_6}$  \\ \hline
Order $1/S$ ($B \neq 0$)   & $\langle \mathcal{\hat{H}}_3 \rangle$ and $\langle \mathcal{\hat{H}}_4 \rangle$ &  $\mathcal{\hat{H}}_3$  \\ \hline
\hline 
\end{tabular}
\caption{Interaction terms retained in the perturbative treatment of the one-magnon dispersion of the SLHAF, without and with an applied magnetic field. Hartree-Fock (HF) and Self-Energy (SE) indicate the types of correction.
\label{tab:interactions}}
\end{table}{}

\subsection{Calculation of the magnon dispersion}

The harmonic magnon dispersion of the quasi-2D SLHAF is given by diagonalizing the quadratic Hamiltonian, $\hat{\mathcal{H}}_2|{\bf k} \rangle=\varepsilon_{\bf k}|{\bf k} \rangle$, which proceeds through the Bogolioubov coefficients $u_{\bf k}$ and $v_{\bf k}$. In our notation 
\begin{eqnarray}
\varepsilon_{\bf k} & = & 2JS (2 + \alpha) \sqrt{(1 + \bar{\gamma}_{\bf k}) (1 - \cos2\theta\,\bar{\gamma}_{\bf k}}), \\
\bar{\gamma}_{\bf k} & = & (\cos k_x + \cos k_y + \alpha \cos k_z)/(2 + \alpha),
\end{eqnarray} 
where ${\bf k} = (k_x,k_y,k_z)$ spans the cubic Brillouin zone and the classical canting angle is given by $\sin \theta = B/B_{\rm sat} = g \mu_{\rm B} \mu_0 B/4JS(2 + \alpha)$.

For all $B \leq B_{\rm sat}$, $1/S$ corrections to the harmonic dispersion include four terms, which we collect as the self-energy
\begin{equation}
\Sigma_{B \neq 0}({\bf k},\omega) = 
    \delta \varepsilon^{\rm HF}_{\bf k} + 
    \delta \varepsilon^{\rm \theta}_{\bf k} +
    \Sigma_{31}({\bf k},\omega) + \Sigma_{32}({\bf k},\omega).
    \label{eq:genericse}
\end{equation}
The first two terms are frequency-independent, corresponding to the Hartree-Fock (mean-field) correction from $\mathcal{\hat{H}}_4$ and the canting-angle renormalization ${\theta} \rightarrow \bar{\theta}$ arising from $\mathcal{\hat{H}}_3$, while the other two terms are frequency-dependent and correspond to one-loop (one-bubble) magnon decay and source processes. As noted in the previous subsection, these can be contrasted with the corrections in zero field, which start at order $1/S^2$ and take the form $\Sigma_{B = 0}({\bf k},\omega) = \delta \varepsilon^{\rm HF}_{\bf k} + \Sigma_{41}({\bf k},\omega) + \Sigma_{42}({\bf k},\omega)$, the first term being the Hartree-Fock correction from $\mathcal{\hat{H}}_4$ and $\mathcal{\hat{H}}_6$ and the other two being self-energy corrections from one-loop (two-bubble) magnon decay and source terms in $\mathcal{\hat{H}}_4$ (Table \ref{tab:interactions}).\cite{Zhito2013}

Next we compare two different approaches that have been adopted to calculate the magnon dispersion from the generic self-energy of Eq.~\eqref{eq:genericse}. The first calculates the corrected dispersion directly ``on-shell'' by inserting the harmonic magnon dispersion into the one-loop diagrams, giving
\begin{equation}
    \bar{\varepsilon}_{\bf k} = \varepsilon_{\bf k} + \delta \varepsilon^{\rm HF}_{\bf k} + 
    \delta \varepsilon^{\rm \theta}_{\bf k} +
    \Sigma_{31}({\bf k},\varepsilon_{\bf k}) + \Sigma_{32}({\bf k},\varepsilon_{\bf k}). \label{eq:onshellse}
\end{equation}
The second calculates the corrected dispersion self-consistently ``off-shell'' as a solution of the Dyson equation, which yields
\begin{equation}
    {\varepsilon}^\ast_{\bf k} = \varepsilon_{\bf k} + \delta \varepsilon^{\rm HF}_{\bf k} + 
    \delta \varepsilon^{\rm \theta}_{\bf k} +
    \Sigma_{31}({\bf k},\varepsilon_{\bf k}^\ast) + \Sigma_{32}({\bf k}, \varepsilon_{\bf k}^\ast).\label{eq:offshellse}
\end{equation}
Both $\bar{\varepsilon}_{\bf k}$ and ${\varepsilon}^\ast_{\bf k}$ can acquire an imaginary part due to spontaneous decays, which adds a layer of complexity in solving the Dyson equation.\cite{Chernyshev2009} In the present case, however, these decay processes are not allowed, which simplifies the calculation.

Strictly speaking, the on-shell approach is restricted to order $1/S$;\cite{Zhitomirsky1999} however, at high fields ($B \gtrsim 0.76 B_{\rm sat}$) it overestimates the decay of the one-magnon branch due to van Hove singularities in the two-magnon continuum once the former enters the latter. The off-shell approach regularizes this deficiency, at the expense of being more costly to calculate and breaking the consistency of the perturbative expansion, in that only some of the terms at orders higher than $1/S$ are included, which can violate the Goldstone theorem. However, at low fields the two dispersion relations $\bar{\varepsilon}_{\bf k}$ and ${\varepsilon}^\ast_{\bf k}$ are almost indistinguishable;\cite{Zhitomirsky1999} because the maximum field in our experiments was only $B/B_{\rm sat} \simeq 0.4$, well below the field-induced decay threshold, for our present analysis we compute the renormalized one-magnon dispersion on-shell ($\bar{\varepsilon}_{\bf k}$), and this is the meaning of our terminology $\overline{1/S}$-SWT. 

\subsection{Calculation of the DSSF}

To obtain the DSSF corresponding to the renormalized single magnons, we first calculate its components in the local frame\cite{Fuhrman2012} by introducing the one-magnon Green functions
\begin{eqnarray}
G^0({\bf k},\omega) & = & \frac{1}{\omega - \varepsilon_{\bf k} + i \delta},  \\
\bar{G}({\bf k},\omega) & = & \frac{1}{\omega - \bar{\varepsilon}_{\bf k} + i \delta}, \label{eq:green} \\
G^\ast ({\bf k},\omega) & = & \frac{1}{\omega - {\varepsilon}^\ast_{\bf k} + i \delta}, \\
G({\bf k},\omega) & = & \frac{1}{\omega - \varepsilon_{\bf k} - \Sigma({\bf k},\omega)}, \label{eq:greensidebands}
\end{eqnarray}
where $\delta$ indicates a small positive number. Here $G^0$ corresponds to the case of harmonic magnons, $\bar{G}$ and $G^\ast$ are the single-pole response functions calculated at the on-shell \eqref{eq:onshellse} and off-shell magnon energies \eqref{eq:offshellse} and $G$ contains the full frequency-dependence of the magnon propagator in Eq.~\eqref{eq:genericse}, for which we use the terminology ``one-magnon sidebands'' (below). As discussed above, the small canting angle at low fields should make $\bar{G}$ \eqref{eq:green} the appropriate choice for analyzing our INS data, but we will show below that $G$ \eqref{eq:greensidebands} is required to capture some of the properties of the LSM.

Within the $\overline{1/S}$-SWT framework, we calculate the one-magnon spectral function as
\begin{equation}
\bar{A}({\bf k},\omega) = -\frac{1}{\pi} {\rm Im} \bar{G}({\bf k},\omega),
\label{eq:spectral}
\end{equation}
which yields the transverse components of the DSSF in the local frame as
\begin{eqnarray}
\bar{\mathcal{S}}^{{x}{x}}({\bf k},\omega) & = & \pi S \Lambda_+^2 (u_{\bf k} + v_{\bf k})^2 \bar{A}({\bf k},\omega), \nonumber \\
\bar{\mathcal{S}}^{{y}{y}}({\bf k},\omega) & = & \pi S \Lambda_-^2 (u_{\bf k} - v_{\bf k})^2 \bar{A}({\bf k},\omega), \label{eq:transverse}
\end{eqnarray}
where $\Lambda_\pm$ are intensity coefficients calculated for each applied field ($B$) and interplane coupling ($\alpha$) from the numerical value of the Hartree-Fock averages.\cite{Fuhrman2012} Technically, these moment-reduction factors, by which transverse spectral weight is transferred to the longitudinal channel by quantum fluctuations, also extend beyond order $1/S$, but are straightforward to include. In addition to the $\Lambda_\pm$ factors, the coefficients $u_{\bf k}$ and $v_{\bf k}$ are also calculated from the harmonic magnon dispersion to maintain the sum-rule on the trace of the DSSF. For simplicity, we proceed to the laboratory frame [Eq.~\eqref{eq:Diagonal-DSSF}] using the unrenormalized canting angle $\theta$ (rather than $\bar{\theta}$, which is almost identical at low fields). 

Turning to the longitudinal component of the DSSF, in the local frame this is dominated by two-magnon excitations, which interact in principle through a one-loop process in $\hat{\mathcal{H}}_4$. However this process is of order $1/S^2$, and hence lies beyond our calculation for the one-magnon dispersion. At order $1/S$, the longitudinal DSSF is given rigorously by the density of states of two non-interacting magnons,
\begin{eqnarray}
\mathcal{S}_0^{{z}{z}}({\bf k},\omega) = & & \frac{1}{2N} \displaystyle{\sum_{{\bf p} \in {\rm BZ}}} (u_{\bf p} v_{\bf p-k} + v_{\bf p} u_{\bf p-k})^2 \times \\
    & & {\rm Im} \displaystyle{\int} \frac{{\rm d}\omega^\prime}{2\pi i} G^0({\bf p},\omega^{\prime}) G^0({\bf k-p},\omega - \omega^{\prime})  \nonumber\\
    = \frac{1}{2N} \displaystyle{\sum_{{\bf p} \in {\rm BZ}}} & & \!\!\!\!\!\!\!\! (u_{\bf p} v_{\bf p-k} + v_{\bf p} u_{\bf p-k})^2
    \delta(\omega - \varepsilon_{\bf p} - \varepsilon_{\bf k-p}). \nonumber
\end{eqnarray}
Although formally correct, this expression has the disadvantage that the transverse and longitudinal response functions are calculated with magnon dispersions of different accuracies, which can become important when comparing their features as functions of energy. This deficiency is straightforward to correct, by calculating the two-magnon density of states in the longitudinal component using the renormalized one-magnon dispersions, which at the on-shell level yields
\begin{equation}
\bar{\mathcal{S}}^{{z}{z}}({\bf k},\omega) = \frac{1}{2N} \! \displaystyle{\sum_{{\bf p} \in {\rm BZ}}} \! (u_{\bf p} v_{\bf p-k} + v_{\bf p} u_{\bf p-k})^2
    \delta(\omega - \bar{\varepsilon}_{\bf p} - \bar{\varepsilon}_{\bf k-p}). \label{eq:longitudinal}
\end{equation}

\begin{figure*}[t]
\includegraphics[width=0.8\textwidth]{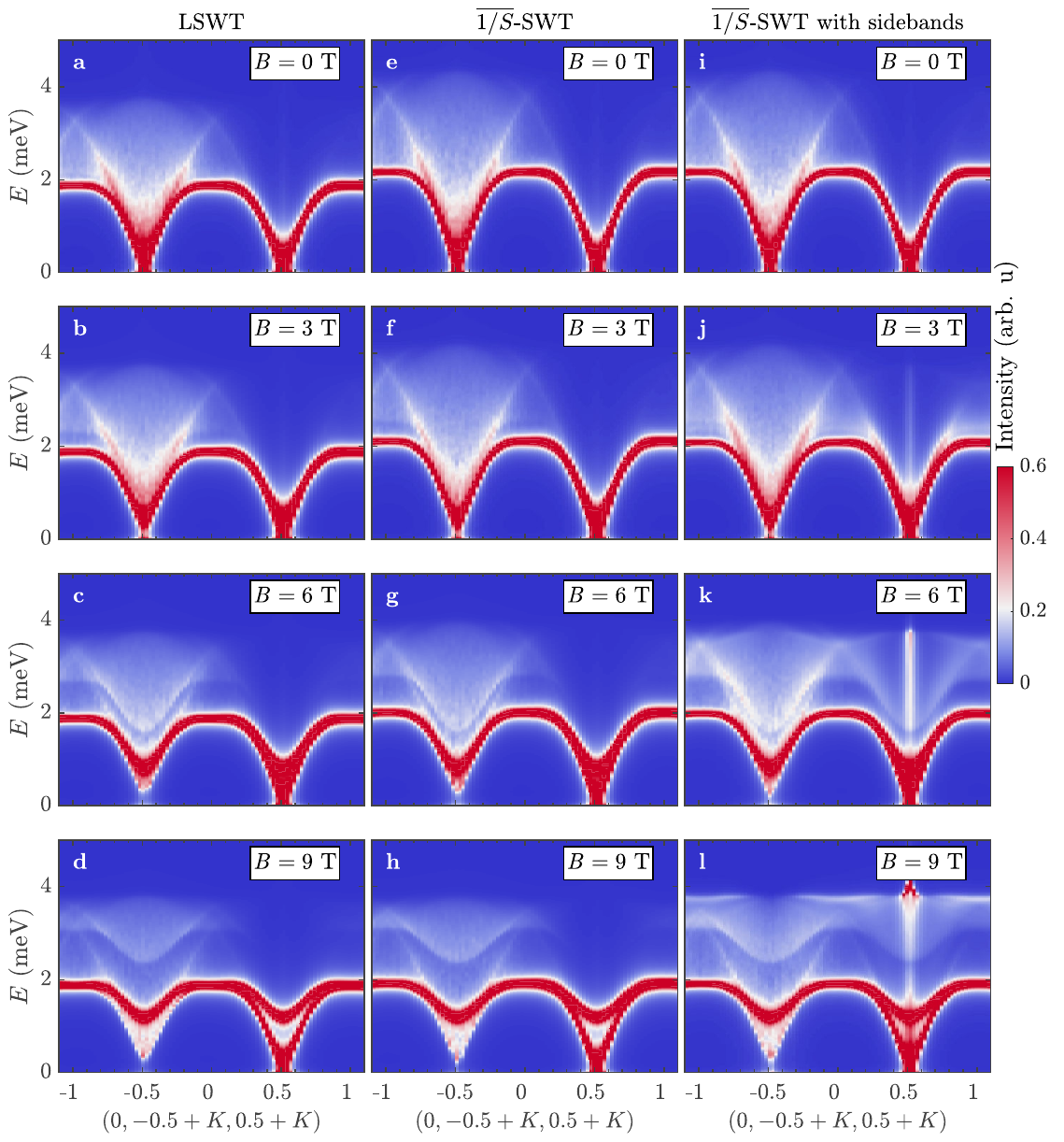}
\caption{{\bf DSSF along MXM from $\overline{1/S}$-SWT.} DSSFs corresponding to momentum transfer $(0, -0.5+K, 0.5+K)$ in the LET experiment, shown for all four applied magnetic fields. {\bf a}-{\bf d} Linear SWT including two-magnon states calculated using the harmonic magnon dispersion. {\bf e}-{\bf h} $\overline{1/S}$-SWT, with both one- and two-magnon excitations calculated on-shell. {\bf i}-{\bf l} $\overline{1/S}$-SWT with additional sidebands allowed for the one-magnon excitations; the vertical line of intensity around $(0, 0, 1)$ is an artifact of the off-shell calculation that is also present, but masked, in Ref.~\onlinecite{Fuhrman2012}} 
\label{fig:SWT-MethodPlots}
\end{figure*}

\begin{figure*}[t]
\includegraphics[width=0.8\textwidth]{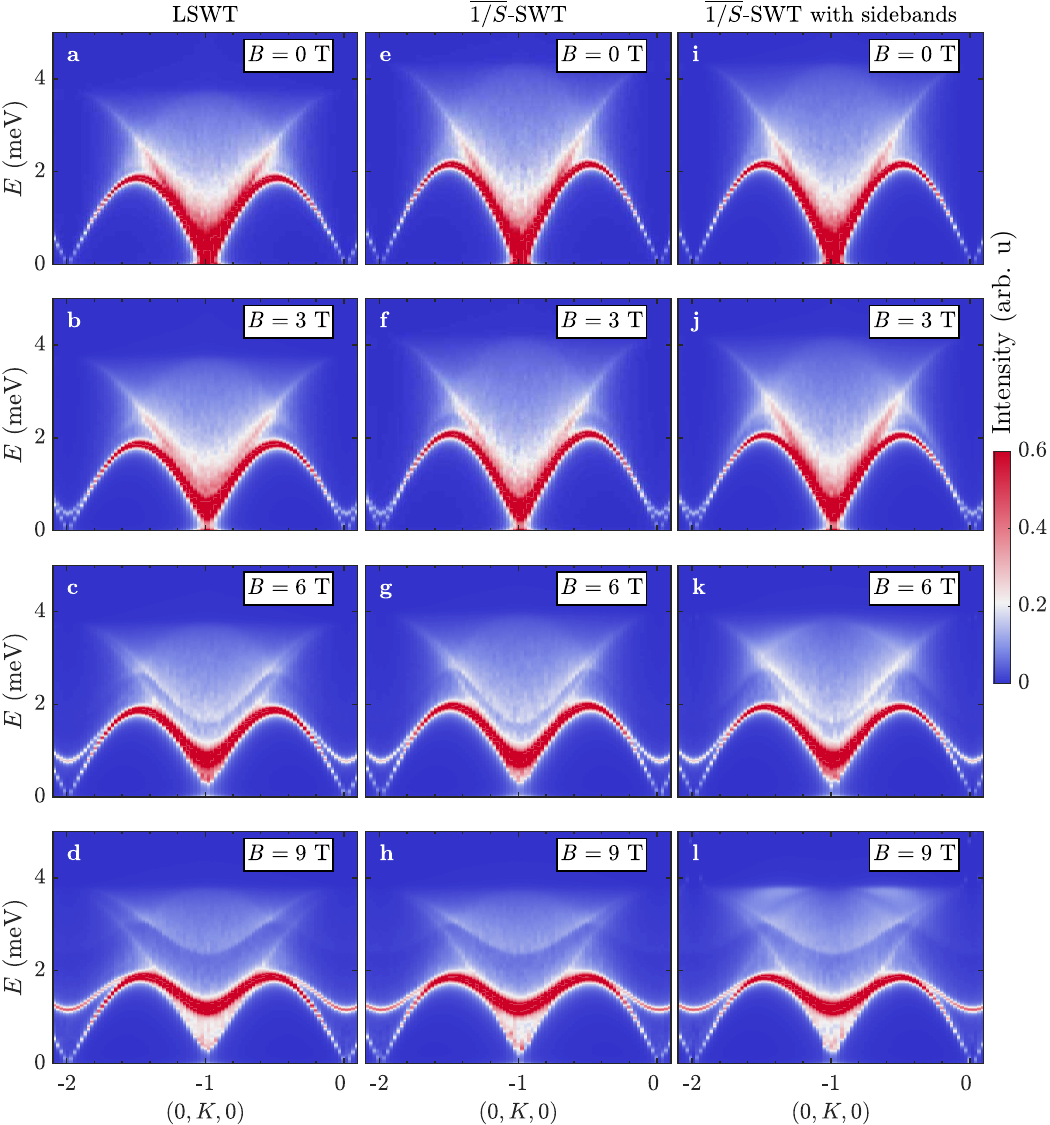}
\caption{{\bf DSSF along $\Gamma$M$\Gamma$ from $\overline{1/S}$-SWT.} As in Fig.~\ref{fig:SWT-MethodPlots} for momentum transfer $(0, K, 0)$. }
\label{fig:SWT-MethodPlots-2}
\end{figure*}

Taken together, an appropriate SWT treatment of our CuF$_2$(D$_2$O)$_2$(pyz) measurements combines the geometrical considerations of Eqs.~\eqref{eq:Diagonal-DSSF} and \eqref{eq:neutron} with consistent calculations of the transverse and longitudinal DSSFs [Eqs.~\eqref{eq:transverse} and \eqref{eq:longitudinal}] that both use the on-shell magnon dispersion ($\bar{\varepsilon}_{\bf k}$) at order $1/S$ [Eq.~\eqref{eq:onshellse}]. Because this approach goes beyond order $1/S$ only formally, but not in execution, we refer to it as a ``calculation at order $\overline{1/S}$.'' As already noted, we will find that the inclusion of one-magnon sidebands in $\mathcal{S}^{{x}{x}}$ and $\mathcal{S}^{{y}{y}}$ captures some of the features of our MPS calculations, and we refer to this as a ``calculation at order $\overline{1/S}$ with one-magnon sidebands.'' 

In SWT, continuous high-energy contributions (sidebands) within the transverse channels reflect the off-shell transfer of spectral weight from the one-magnon pole to the two-magnon continuum by cubic vertices. This effect can be captured by replacing $\bar{A}({\bf k},\omega)$ with ${A}({\bf k},\omega) = -1/\pi\, {\rm Im}\, {G}({\bf k},\omega)$ in Eq.~\eqref{eq:transverse} so that the transverse components of the DSSF in the local frame become
\begin{eqnarray}
\bar{\mathcal{S}}_{\rm sb}^{{x}{x}}({\bf k},\omega) &=& - S \Lambda_+ (u_{\bf k} + v_{\bf k})^2 {\rm Im}\left[\frac{1}{\omega - {\varepsilon}_{\bf k} - \Sigma({\bf k},\omega )}\right], \nonumber \\
\bar{\mathcal{S}}_{\rm sb}^{{y}{y}}({\bf k},\omega) &=& - S \Lambda_- (u_{\bf k} - v_{\bf k})^2 {\rm Im}\left[\frac{1}{\omega - {\varepsilon}_{\bf k} - \Sigma({\bf k},\omega)}\right]. \nonumber  \\ \label{eq:transverse-sidebands}
\end{eqnarray}
This transfer also affects the longitudinal component, which we describe by introducing a quasiparticle-residue formulation for the one-magnon Green function,\cite{Chernyshev2009}
\begin{equation}
G({\bf k},\omega) \approx \frac{Z_{\bf k}}{\omega - \bar{{\varepsilon}}_{\bf k} + i\delta} + G_{\rm inc}({\bf k},\omega) \label{eq:greenresidue} 
\end{equation}
with
\begin{equation}
Z_{\bf k} = {\rm Re} \left[1 - \frac{\partial \Sigma({\bf k},\omega)}{\partial \omega}\right]_{\omega = \bar{\varepsilon}_{\bf k} }^{-1}.
\end{equation}
The advantage of this formulation is that the quasiparticle pole can be inserted directly into the longitudinal DSSF,
\begin{eqnarray}
\bar{\mathcal{S}}_{\rm sb}^{{z}{z}}({\bf k},\omega) & = &  \frac{1}{2N} \displaystyle{\sum_{{\bf p} \in {\rm BZ}}} 
    Z_{\bf p} Z_{\bf k-p} (u_{\bf p} v_{\bf k-p} + v_{\bf p} u_{\bf k-p})^2 \nonumber \\
    & & \qquad\qquad \times \; \delta(\omega - \bar{{\varepsilon}}_{\bf p} - \bar{\varepsilon}_{\bf k-p}), 
\end{eqnarray}
at the expense of neglecting the incoherent sidebands ($G_{\rm inc}({\bf k},\omega)$ in Eq.~\eqref{eq:greenresidue}). 

\subsection{Calculation and results}

We computed the DSSF within $\overline{1/S}$-SWT by using and improving the Matlab code developed in Ref.~\cite{Fuhrman2012} for the quasi-2D SLHAF. In all of these calculations, Hartree-Fock averages for a given $B$ and $\alpha$ were computed using Simpson's Rule on a $1000^3$ grid to obtain six-digit accuracy. Unless otherwise noted, all self-energy calculations were performed using the Monte Carlo method with $N_{\rm MC} \geq 10^5$ and Brillouin-zone integrals for the two-magnon density of states with $N_{\rm 2M} \geq 10^4$. In the following we use a small Lorentzian broadening of $\delta = 0.05$~meV in our colour contour figures, $\delta = 0.03$~meV for line cuts that compare $\overline{1/S}$-SWT with MPS and INS and $\delta = 0.005$~meV to study the positions of van Hove singularities.

\begin{figure*}[t]
\includegraphics[width=0.99\textwidth]{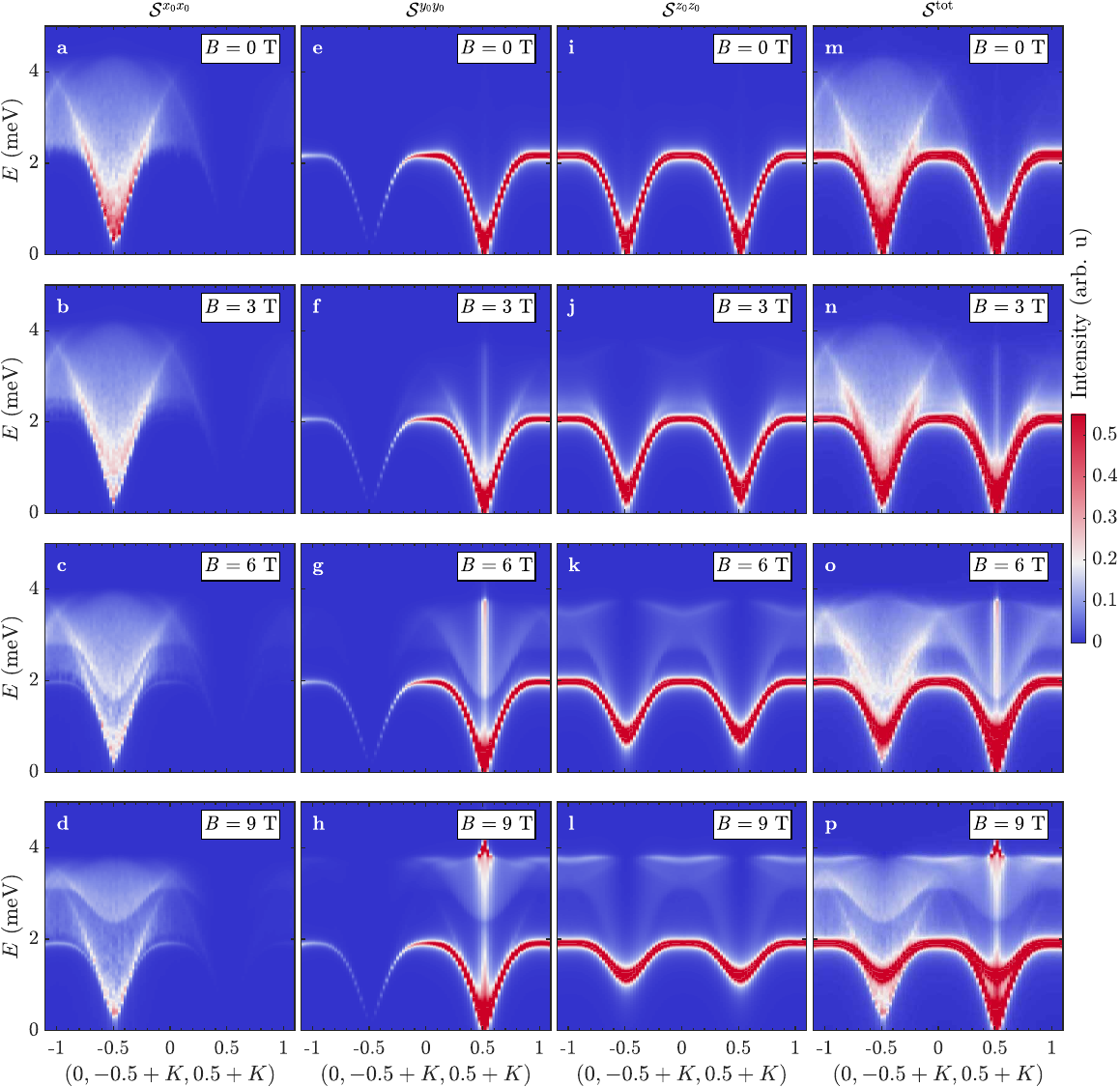}
\caption{{\bf Channel-resolved DSSF along MXM from $\overline{1/S}$-SWT.} DSSF components $\mathcal{S}^{x_0x_0}$ (a-d), $\mathcal{S}^{y_0y_0}$ (e-h), $\mathcal{S}^{z_0z_0}$ (i-l) and their sum, denoted $\mathcal{S}_{\rm tot}$ (m-p), shown at the four magnetic fields of the LET experiment for momentum transfer $(0, -0.5+K, 0.5+K)$, calculated by $\overline{1/S}$-SWT with additional sidebands allowed for the one-magnon excitations.}
\label{fig:SWT-Channels}
\end{figure*}

\begin{figure*}[t]
\includegraphics[width=0.99\textwidth]{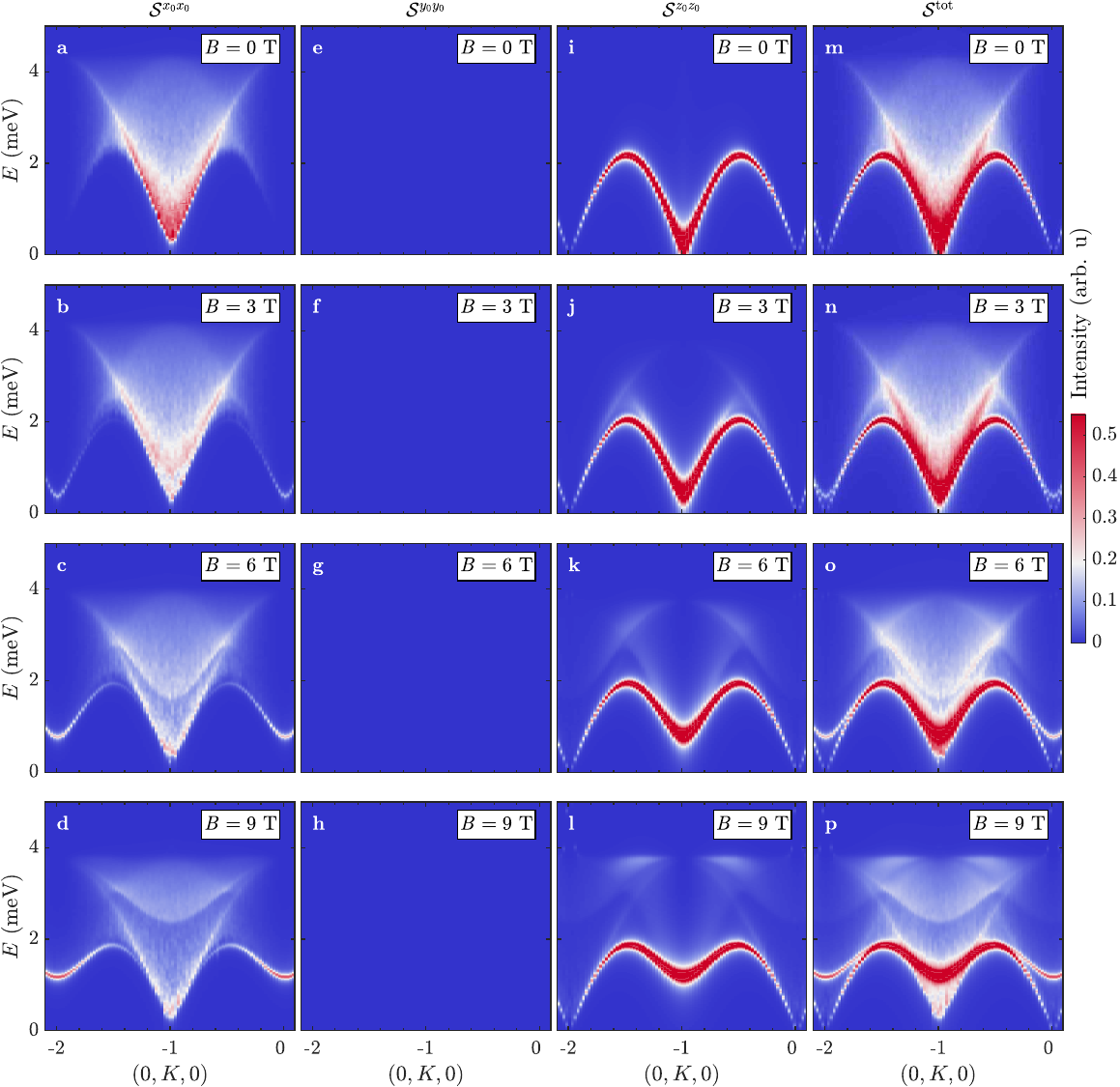}
\caption{{\bf Channel-resolved DSSF along $\Gamma$M$\Gamma$ from $\overline{1/S}$-SWT.} As in Fig.~\ref{fig:SWT-Channels} for momentum transfer $(0, K, 0)$.} 
\label{fig:SWT-Channels-2}
\end{figure*}

The results of these calculations at the different levels of approximation are presented in Figs.~\ref{fig:SWT-MethodPlots} and \ref{fig:SWT-MethodPlots-2} for the momentum-energy scans shown in Fig.~2 of the main text. Other than the artifact of the off-shell calculation appearing around the centre of the magnetic Brillouin zone, the $\overline{1/S}$ calculation with sidebands captures more features of the INS and MPS data, and thus we adopt this level of approximation in the discussion to follow. SWT is particularly valuable for interpreting the DSSFs of the different spin channels, and hence in Figs.~\ref{fig:SWT-Channels} and \ref{fig:SWT-Channels-2} we present the channel-resolved response functions for a field corresponding to $B = 9$~T, with the polarization factor and canting-angle corrections applied. These calculations correspond to the MPS results shown in Figs.~\ref{fig:MPS_Pol_Direction1} and Fig.~\ref{fig:MPS_Pol_Direction2}.

\begin{figure}[t]
\includegraphics[width=0.98\columnwidth]{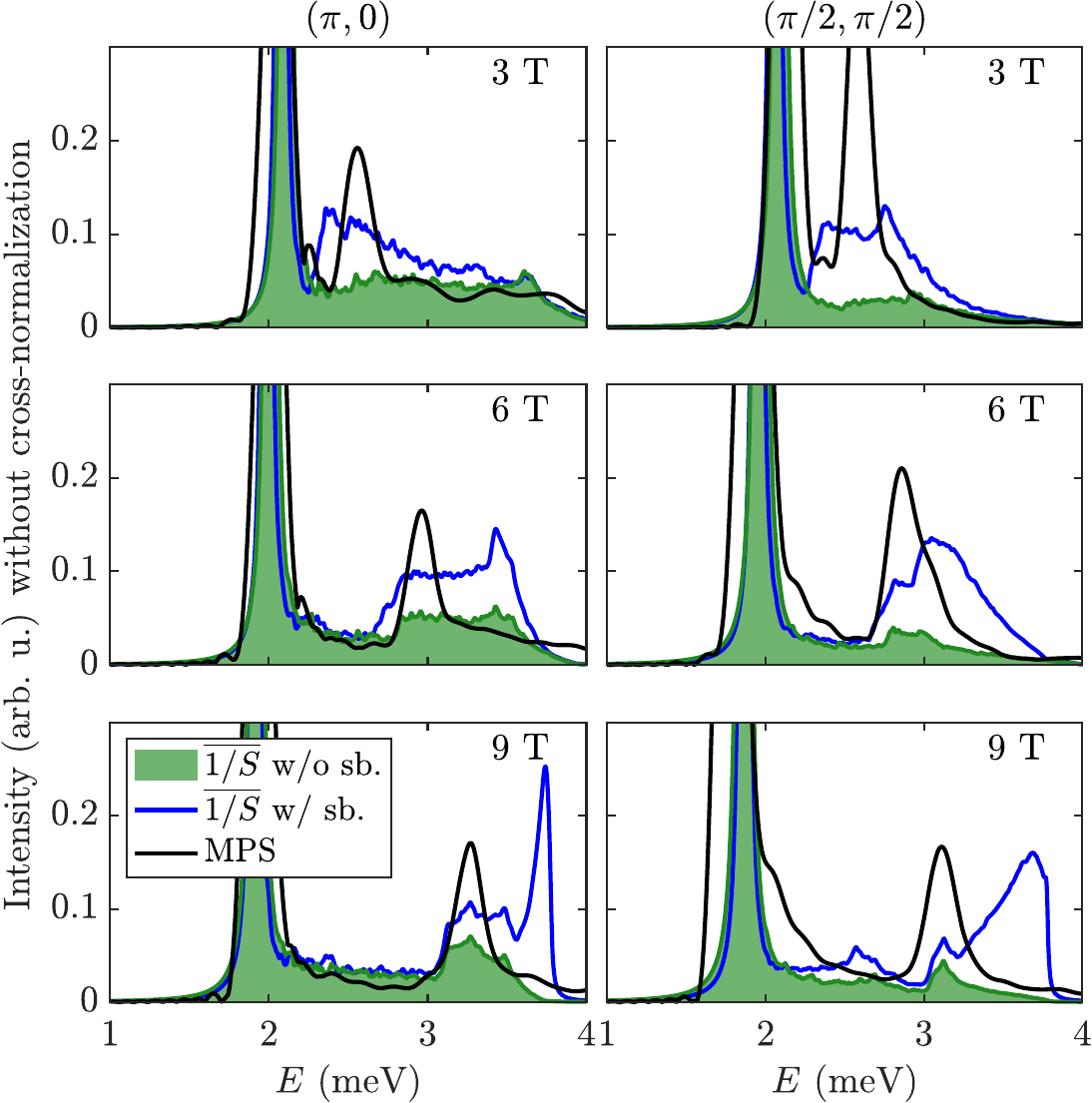}
\caption{{\bf Van Hove singularities in the two-magnon density of states.} Comparison between the positions of van Hove singularities in the SWT continuum and the LSM predicted by MPS.}
\label{fig:SWT-vanHove}
\end{figure}

\begin{figure}[t]
\includegraphics[width=0.99\columnwidth]{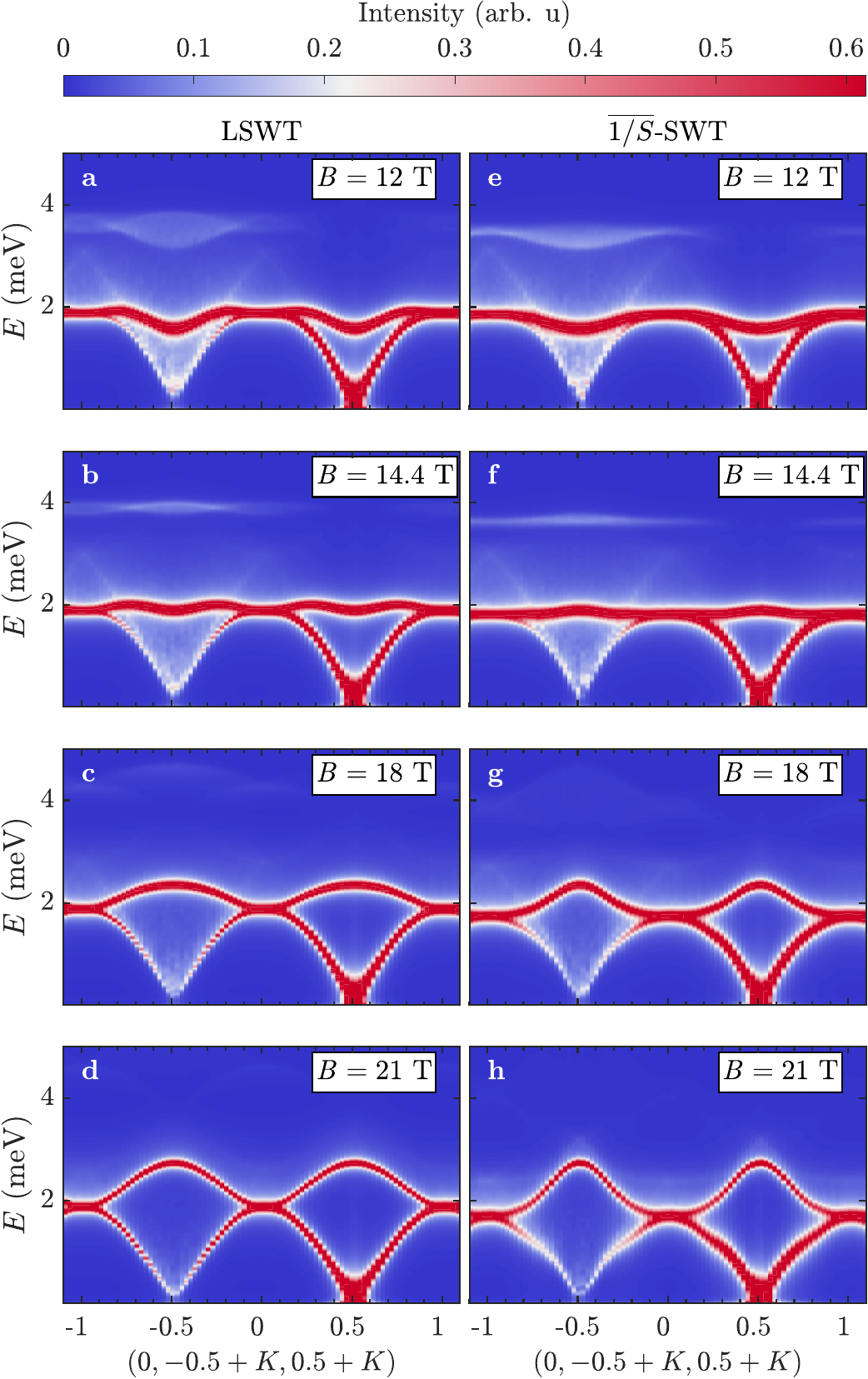}
\caption{{\bf High-field DSSF along MXM from $\overline{1/S}$-SWT.} DSSFs corresponding to momentum transfer $(0, -0.5+K, 0.5+K)$ for fields in the vicinity of half saturation, which are beyond the range of our experiment. {\bf a}-{\bf d} Linear SWT including two-magnon states calculated using the harmonic magnon dispersion. {\bf e}-{\bf h} $\overline{1/S}$-SWT with both one- and two-magnon excitations calculated on-shell.}
\label{fig:SWT-HalfSat}
\end{figure}

\subsection{Two-magnon resonance in the continuum}

It is clear from Figs.~\ref{fig:SWT-MethodPlots}-\ref{fig:SWT-Channels-2} that SWT at this level does not capture the LSM we observe by INS and MPS. As in Figs.~4d and 4h of the main text, the two-magnon continuum within $\overline{1/S}$-SWT remains broadly distributed in energy at all wavevectors. This reflects the fact that $\overline{1/S}$-SWT includes magnon interaction effects only at the level of a self-energy of individual magnons, and does not contain direct magnon-magnon interactions in the calculation of the continuum. In this way our $\overline{1/S}$-SWT calculation allows us to benchmark the type of physical processes that give rise to a resonance as sharp as the LSM. As noted above, the effects of these magnon-magnon interactions have been captured by a numerical high-order expansion,\cite{Powalski,Powalski2018} although to date this has been performed only at zero field and not in a state with canted magnetic order. 

However, our $\overline{1/S}$–SWT calculations show that including sidebands (Figs.~\ref{fig:SWT-MethodPlots}i-l and \ref{fig:SWT-MethodPlots-2}i–l) does produce a continuum edge across the full momentum range, which matches the location of the LSM. Without sideband effects, the continuum arises only from the longitudinal ${x_0x_0}$ and ${z_0z_0}$ channels, and is therefore suppressed very strongly at $(0,0,1)$ and $(-2,0,0)$ by polarization factors and the small canting angle, leaving it visible only near $(0,1,0)$ (Figs.~\ref{fig:SWT-MethodPlots}e-h and \ref{fig:SWT-MethodPlots-2}e-h). Sidebands transfer spectral weight into the transverse ${y_0y_0}$ channel, which is not suppressed at $(0,0,1)$, thereby restoring a continuum edge across  the full {\bf Q} range. At $(-2,0,0)$ this transverse channel is again suppressed, as observed in experiment. Physically, sidebands redistribute weight from the one‑magnon mode into the two‑magnon continuum in a way that produces a more isotropic spin response than bare $\overline{1/S}$‑SWT can capture.

\begin{table}
\begin{tabular}{|ll|l|l|} \hline\hline
${\bf p}$ & ${\bf q}$ & $f({\bf p},{{\bf q}})$ & Singularity Type \\ \hline
L & ZB & $\varepsilon_{\rm ZB} + \Delta + \alpha (\tilde{p}_x^2 + \tilde{p}_y^2) - \beta_x \tilde{q}_x^2 - \beta_y \tilde{q}_y^2$ & Logarithmic \\
L & G & $\Delta + \alpha (\tilde{p}_x^2 + \tilde{p}_y^2) + c |\tilde{\bf{q}}|$ & Power-law  \\
L & L    & $2 \Delta + \alpha (\tilde{p}_x^2 + \tilde{q}_x^2 + \tilde{p}_y^2 + \tilde{q}_y^2) $ & Step \\ \hline
\end{tabular}
\caption{Nature of van Hove singularities involving the Larmor mode in the two-magnon density of states, $\mathcal{D}_2({\bf k},\omega)$, in the low-field regime as a function of the independent momenta ${\bf p}$ and ${\bf q}$ contributing to the density of states. L denotes Larmor, ZB zone-boundary and G Goldstone. 
\label{tab:van-hove}}
\end{table}{}

Because this edge feature in the continuum tracks the position of the LSM, we presume that it can be considered as providing a basis for the strong bootstrapping effect of magnon-magnon interactions that results in the appearance of a sharp resonance within the continuum. To trace the origin of this feature, we consider the presence and nature of van Hove singularities in the two-magnon density of states of the SLHAF. We proceed for simplicity by using harmonic magnons, but this aspect of the analysis is the same for the renormalized dispersions $\bar{\varepsilon}_{\bf k}$ and $\varepsilon^\ast_{\bf k}$. At the low magnetic fields we consider, the DSSF is dominated by three regions of the Brillouin zone where the dispersion is either nearly massless or disperses quadratically away from a flat region around an extremum. Denoting small wavevectors away from the extrema by $\tilde{\bf{k}}$, these are: (i) the Larmor mode near ${\bf k} = 0$ which is a local minimum ($\alpha > 0$) 
\begin{equation}
    \varepsilon^{\rm L}_{\bf k} \approx \Delta + \alpha (\tilde{k}_x^2 + \tilde{k}_y^2), 
\end{equation}
where $\Delta = g \mu_B B$ is unaffected by quantum corrections due to the Larmor theorem; (ii) the zone-boundary excitations extending from ${\bf k} = (\pi,0)$ to $(\pi/2,\pi/2)$, 
\begin{equation}
    \varepsilon^{\rm ZB}_{\bf k} \approx \varepsilon_{\rm ZB} - \beta_x \tilde{k}_x^2 - \beta_y \tilde{k}_y^2, \,\, {\rm with}  \,\,\,\, \varepsilon_{\rm ZB} = 4 J S,
\end{equation}
which present a global maximum ($\beta_{x,y} > 0$ at low fields); (iii) the Goldstone mode near ${\bf k} = (\pi,\pi)$,
\begin{equation}
    \varepsilon^{\rm G}_{\bf k} \approx  c |\tilde{\bf{k}}|, \,\, {\rm with}  \,\,\,\, c = 2\sqrt{2} J S,
\end{equation}
which is a global minimum. 

The two magnon density of states,
\begin{equation}
\mathcal{D}_2({\bf k},\omega) = \frac{1}{N} \displaystyle{\sum_{{\bf p} \in {\rm BZ}}} \delta(\omega - {\varepsilon}_{\bf p} - {\varepsilon}_{{\bf q} = {\bf k-p}}), 
\end{equation}
displays van Hove singularities when the function $f({\bf p},{{\bf q}})={\varepsilon}_{\bf p} + {\varepsilon}_{{\bf q} = {\bf k-p}}$ has extrema. In general, local minima and maxima give rise to step singularities, while saddle points produce logarithmic divergences, which are more pronounced. In Table \ref{tab:van-hove} we follow Ref.~\cite{Mourigal2010} to classify the three types of singularity involving the Larmor mode, in order of decreasing strength, as functions of the independent momenta ${\bf p}$ and ${\bf q} \equiv {\bf k} - {\bf p}$. All of these singularities  contribute at least a step discontinuity to the continuum edge that tracks the position of the LSM, with the strongest singularity (logarithmic) originating from the combination of a Larmor mode with zone-boundary magnons. In this way even linear SWT provides additional insight into the physics underlying the emergence of a non-perturbative resonance in the continuum as strong as the Larmor-shadow mode. In Fig.~\ref{fig:SWT-vanHove}, we compare the spectral functions calculated by $\overline{1/S}$–SWT with sidebands to our MPS results for the two zone-boundary wavevectors of Fig.~5 at 3, 6 and 9~T, which enables us to relate the positions of the van Hove singularities in SWT with the actual position of the LSM.  

Finally, in Fig.~\ref{fig:SWT-HalfSat} we use our $\overline{1/S}$-SWT framework to compute the spectral function at applied magnetic fields around and above half saturation ($B_{\rm sat}/2 = 14.4$~T). At this field the Larmor mode in linear SWT transforms in from a local minimum to a global maximum of the magnon dispersion, and this change is clearly associated with the disappearance of any enhanced density-of-states effects in the two-magnon continuum. As noted in the main text, we expect on this basis that the LSM ceases to exist for fields above half saturation, which would be consistent with the collapse of the shadow feature observed in the honeycomb-lattice Heisenberg antiferromagnet.\cite{hlsm2025}

\end{document}